\documentstyle[12pt, fleqn,epsf,epsfig]{article}

\title{Functional Dynamics I:\\Articulation Process }
\author{
Naoto Kataoka and Kunihiko Kaneko\\
  {\small \sl Department of Pure and Applied Sciences}\\
  {\small \sl University of Tokyo, Komaba, Meguro-ku, Tokyo 153, JAPAN}\\
}

\date{}
\begin{document}
\maketitle
\begin{center}
submitted to Physica D
\end{center}

\abstract

The articulation process of dynamical networks is studied with
a functional map, a minimal model for the dynamic change of relationships
through iteration.  The model is a dynamical system of a function $f$,
not of variables, having a self-reference term $f \circ f$,
introduced by recalling that
operation in a biological system is often applied to itself, as is
typically seen in rules in the natural language or genes.
Starting from an inarticulate network,
two types of fixed points are formed as
an invariant structure with iterations.
The function is folded with time, until it has finite or infinite
piecewise-flat segments of fixed points, regarded as articulation.
For an initial logistic map, attracted functions are classified
into step, folded step, fractal, and random phases,
according to the degree of folding.
Oscillatory dynamics are also found, where function values are mapped to several
fixed points periodically.
The significance of our results to prototype categorization in language
is discussed.

\section{General Introduction}

In studying a biological system,
we face the problem of how rules are generated.
In simulations of physical systems, a rule is given in advance from a
natural law.  On the other hand, in trying to model
a biological system at a biological level, we need
to study the origin or dynamics of the rule itself.

There are two possibilities for such a study.
In one approach, one starts from a level that is more microscopic than that of the biological system
(e.g., the chemical reaction level) and determines form a description of 
of behavior on this smaller scale how a rule (e.g., a rule for
cell differentiation)
is formed at a biological level.  In the other approach,
one attempts to start at a  biological level from the beginning.

Since the latter approach, when successful, can allow for the extraction of  
the essential features  of a biological system, we consider 
this approach here.
In attempting to employ this approach, however,
we face the following difficulty.  
In this approach, the rule (operator) and variable (operand) are not initially distinguished,
and they should be at the same level in the beginning.
For example, consider a gene. This is nothing but a set of chemicals
within a DNA molecule.  Among the chemicals in a cell, 
the chemicals contained in the DNA molecule constitute some kind of `rules' for other chemicals.
A more straightforward example is seen in the problem of
language, where code and encoding are not distinguished at a descriptive level. 
If a language were nothing but a signal (as is the case for the emergency calls of birds), 
there would be no need to distinguish code from encoding.
However, it is thought that there is something called `encoding' within our language.
In spite of the conviction that there is meaning in an uttered phrase or piece of text, this meaning can
 be described only at the code level.
Although in describing encoding, it is represented by a set of codes,
the encoding in the natural language needs something beyond such a set.
(See Sec.2 for discussion on language.) 

In view of the above considerations, 
it would seem that we need to construct a model in which the rule (operator) is
not initially separated from the variable (operand).
In the case of language, we have to start from a model without distinction between
code and encoding.  Since a rule is not distinguished from
the entity to which the rule is applied, the operation of a rule can be applied to
itself.  Hence, one has to consider `self-reference' seriously.
As will be seen, we consider the dynamics of networks of relationships.  
These dynamics are formulated in terms of the dynamics of function,
instead of dynamics of variables.
In this formulation, self-reference is expressed by the operation of the function 
on itself (i.e., $f \circ f(x)$).

The second problem in a biological system is the emergence of symbols,
made possible by `articulation' of continuous objects.
Here, `articulation' means the categorization of the words or molecules.
Although this categorization is usually used 
as a classification of some elements at the same level into some groups,
the inseparability of the rule and the variable has to be considered,
as pointed out in the discussion of the first problem.
Thus, the articulation is used to describe a categorization of words and rules.
Through the dynamics of networks of relationships of objects,
some object begin to act as a symbolization of other
objects. This can be seen in the emergence of 
`information carrying molecules', such as DNA, and in the emergence of language
with some symbols. Once the symbols are formed, they remain as stable
objects, while connections to such symbols are formed as relatively stable
relationships. In the present paper, the process of articulation
is studied through the dynamics of functions.

The third problem in the study of a biological system is understanding the formulation of a rule to change
the relationships among objects.  Such a rule leads to dynamics of the
symbols, but it also refers to the object assigned to the symbol.
For example, rules for differentiation are formed in reference to genes (symbols) 
in DNA, but these rules also can depend on other chemicals associated with the
genes. Although grammar consists of rules concerning symbols, often these rules are
not completely syntactic, but rather depend on the objects that are assigned to symbols.

The fourth problem is that of hierarchical rule formation, as is seen in
the development process of an organism or in the language.  
Here the dynamics of the changes of the rule itself emerge as a meta-rule.

The fifth problem is understanding the formation of a `social' rule among some subjective, tissues or organs.
The `society' of `units' is assumed to have a common rule. 
On one hand, this rule is given from the form of the whole society, and
on the other hand, it is generated from the concurrence among some units.

In a series of papers, we attempt to construct a mathematical
framework that solves the above five requisites. In this, the first paper of the series,
we consider the basic structure of the functional map, and discuss the articulation process.
The third and fourth problems, which are more central to the present formulation, 
will be discussed in the next paper.
The fifth problem will be discussed in a third paper.

\section{Introduction}

Before discussing our model and the results of
articulation dynamics, we briefly mention our motivation
in connection with the study of language.
Since some epistemological understanding are necessary to study natural language, 
we briefly review trends in the philosophy of language.\footnote{If the reader is mainly 
interested in functional dynamics as a dynamical system, 
one can skip the following arguments and jump to the last six paragraphs of this section.}

Although a logical model for language was extensively 
studied from the 1920s to the 1950s in the context of `logical positivism' \cite{Wais}\cite{Ayer}, 
this approach
could not cover the entire area of cognition or semantics.
Theory-laden nature of measurement \cite{Hanson} and the position of the real world in the theory 
then led to a shift from
`logical positivism' to `logical negativism' \cite{Quine}.
There, it has been recognized
that a theory of natural language is not sufficient without 
the concept of cognition.
By this recognition, new trend in the language theory started, that is 
the theory of the `speech act' (usage of language), 
which focuses on `intention' \cite{Austin}\cite{Searle}.
This theory, however, is grossly insufficient given the diversity of utterances.
In the logic-based theory, 
the problem of how the continuous world can be represented by combinations of finite words is not addressed.
In the theory, only minimal elements and rules governing these elements from the outset are included.

In contrast to the above theories, `structurism' attempts to attack the background of language \cite{Sau}.
It the structurism, one focuses on 
how continuous objects are articulated into words, depending on the structure of all other words.
However, this theory deals only with 'static' structure, and
can only describe a `snapshot' property of the language mechanism,
which indeed has developed from each individual's birth and evolved 
historically since its origin.
There are some philosophical discussions \cite{Deleuze} with regard to replacing
a theory of the static structure of language by a theory including dynamics,
but these remain speculative, without 
any concrete mathematics.
A mathematical model
for epistemology or articulation dynamics is necessary to study 
natural languages.
However, in a mathematical model we cannot deal with cognition directly.
Thus, a `detour' is needed to understand the underlying basic
structure of languages.

Since phones and letters are only signs, it is not possible to
distinguish a code from the encoding entity. 
There is only a circulation of signs.
The code-encoding relation is decided by correspondence to the real 
world, although the dynamics of signs seems to be self-driven 
at the descriptive level, too.

The dependence of the rules imposed on signs upon the description level
is evidenced by the existence of dictionaries.
This existence implies that a sentence cannot be produced only from the 
formal logic.
If all sentences were made logically, dictionaries would be much thinner or perhaps would be unnecessary.
The role of dictionaries shoulder the redundancy in language, which is beyond the 
formal logic.
This redundancy is related to cognition and derives from the uniqueness of our world.
Hence, this redundancy shown by the existence of dictionary has to be considered seriously
for the study of cognition. 
In addition to redundancy, an important 
characteristic of dictionaries is the circulation among words.
A word is described by other words, which are, in turn, described by still other words.
In other words, there are self-referential relations among signs.
This self-referential structure is taken into account for our
`detour' of understanding the basic structure of language systems.

However, if we focus only on the static network of words, 
the detour is not relevant to the study of cognition.
The static network that has already been articulated is only a snapshot of language.
To study the static structure of language, we need a database of
existing languages, which we cannot include in our preliminary study with an
abstract mathematical model.  Rather, in our study, we focus on
some universal structure that the 
articulation in dynamical networks possesses, from which
we regard the articulation
process of cognition as one important aspect of
the natural language.
Of course, this is a difficult problem
that cannot be solved in a single paper. Here we present a
first attempt at solving this problem by introducing an abstract model for
a dynamical network and study a class of general phenomena.

In previous studies of natural language,
it is common for code and encoding to be divided.
The celebrated theory of code is the `generative grammar'
(Chomsky 1955) \cite{Chom}, 
while `cognitive semantics' (Lakoff 1987) \cite{Lakoff} is a theory for the encoding.

For the study of codes, 
generative grammar deals with transformation of words and
sentences.  Given a formal system, such transformations are
classified into several classes \cite{cal}, according to
computation theory.  Study of generative grammar has succeeded in
describing how an already described language is structured.
However, this is just a one-way flow 
from the phones or letters to the language structure.
Natural language cannot be generated only by this language structure.
To produce a sentence, it is inevitably necessary to refer to
the real world.
If we have to refer to real world structure in order to produce sentences,
the syntactic rules must be complemented by additional cognitive rules
in order for a machine to be able to speak or a program to be able to write.
Thus, a theory or a model of articulation of language is necessary, 
for example, to construct
a machine which can use language.

In this respect,
cognitive semantics deals with the structure of the linguistic network 
among words, which is preserved and constructed
through the iteration.
It is proposed that there is semantic structure 
(a network of category, or a set of sets of words),
which is formed through these iterations and is robust to some degree.
This semantic structure is an articulation of the real world.
Although a category is a set of words, it is not decided only from properties of objects,
but also through cognitive processes. 
Because of this property of a category, it is not the case that
all the words in the category have the same status, and thus there is asymmetry 
in the structure of the category derived from cognition.
A reference element which is suitable for the cognitive process 
(restricted physically, socially and so on)
is called a `prototype' \cite{Lakoff}, 
while other elements in the same category can have various status.
The repeated use of language produces a semantic network of category whose prototype
is essential to the articulation in the language.

In  communication it is essential that some entities are identified
through iteration which is restricted, for example, by cognitive,
physical, social and common laws.
Iteration gives the foundation of the language.
If the language were not formed through iteration,
it would be destroyed into fragments like those Borges imagined \cite{Borges} or in Finnegans Wake.
Some iteration processes preserve meaning, while some others
alter existing meanings. With the confliction of these two processes,
one can describe something already described, while one can
also think about what has not yet been described.
By iterating encoded words in a language network,
such a network is formed as a connection of codes.

A fuzzy logic can be one of the tools to extract semantic structure,
because each element which belongs to the category can have a different status.
However, semantic structure formed through the cognitive process has a dynamical aspect, and hence
a fuzzy `logic' is not sufficient to deal with it.
To consider cognitive and semantic processes, a dynamical model of fuzzy logic \cite{Grim} and
a dynamical network of proposition in a fuzzy logic as a functional form \cite{Tsuda} have been proposed.
However, to study the duality between rule and code directly,
we must avoid such a logic-based approach.
Starting from an inarticulate system without logic, the emergence of articulation is studied in this paper, while
the emergence of rule (logic) is studied in the next paper, II.\\

To sum up the long discussion so far, we have to study a system having following features: 
(a) networks of relationships change in time, 
(b) rules (operators) and variables (operands) are not initially separated 
(c) the rules (operators) are applied to themselves, since operands are not separated 
(d) through the iteration of the rules
(e.g., the use of language), some robust structure in the network
is formed, by which the roles of operators (rules) and operands are
separated (e) from 
continuous objects, discrete symbols are formed through the iteration of
the rules to change the network of relationships.

To represent an inarticulate network mathematically,
we adopt a function (network) instead of a variable,  
as its minimal element (word).  This function $x \rightarrow f(x)$
can represent the network of
relationship, or a filter from inputs to outputs.  With this representation by
a function, the application of a rule to itself is represented by
the self-reference term $f \circ f(x)$ so that the relationship 
$x \rightarrow f(x)$ is applied to $f(x)$ itself.  Since the operand
of the function is the function itself in this term, the operator
and operands are not separated.

In addition to the robust structure with respect to iteration,
the language has ability to create variety.  Objects indicated
by codes can vary in context and in time.
Also, if a novel object, which has not existed before, appears,
language is able to refer to it. 
Conversely, we can describe only what our language can approach.
In spite of this restriction, we can always face  
new objects in a new manner.
Hence the network structure in language has variability
to support such diversity.  To cope with such variability,
a dynamics of the function $f(x)$ will be introduced into our model,
so that the relationship $x \rightarrow f(x)$ can change dynamically.
We study the evolution of the function $f_n(x)$ at time step $n$, following
functional dynamics depending on $f_n(x)$ and $f_n \circ f_n(x)$.

By taking a continuous variable $x$, the problem of 
articulation will be studied
as a classification process how $f_n(x)$ converges to distinct intervals
in each of which $f_n(x)$ takes a different constant value. 
For a given value $a = f_n(x)$ the inverse set $I_n = f^{-1}_n(a)$ is given
as an articulated class.
This means that the filter articulates the continuous world $x$
into some segments according to the value $f_n(x)$.

This paper is organized as follows.
We propose a model of the articulation process in  Sec.3, where
a map for a function $f(x)$, not for a variable $x$, is introduced.
The dynamics of this function are given by a balance between the original map
$f(x)$ and its iteration $f \circ f(x)$.
As a preliminary step for later studies, we discuss some
elementary properties of these functional dynamics in Secs.4 and 5.
In Sec.4, we discuss the case in which the dynamics are given only by 
$f \circ f(x)$, to clarify  periodic dynamics of the network.
In Sec.5, we discuss the simplest case with a monotonic function,
to understand the minimal articulation property of our dynamics.
In Sec.6 we choose a single-humped map as an initial function,
to understand the articulation process 
balancing between self-reference and diversity.
The limiting forms of the functions are classified in Secs.7 and 8
by introducing several quantities characterizing stepwise, fractal and
other singular functions. A self-folding mechanism is discussed in detail.
Periodic solutions in this model are given in Sec.9. 
Summary and discussion are  given in Secs.10 and 11.

The possible class of behavior in our functional dynamics is not restricted to those discussed
in the present paper.
Here, we discuss only periodic structures generated by isolated fixed points.
Indeed, a method to determine periodic points mapped to continuous fixed points is presented in a subsequent
paper, where hierarchical rules to change rules will be organized in our functional dynamics \cite{II}.

\section{The Model}

As a minimal model of the dynamics of articulation or categorization, we introduce
a functional dynamics. Here a function
represents a relation among elements that can be words, chemicals, and so forth.
In this abstract model, a function
corresponds to a network consisting of directed elements.
We study some characteristic features of functional dynamics
with the iterated application of the function to itself.

Given an initial network, it evolves
according to a transformation rule that is determined by the shape of the network itself.
On one hand, the function is postulated to have
a self-referential property through the operation of the function on
itself. On the other hand, the function
is required to have the ability to exhibit diverse behavior, 
to drive the network to include a variety of elements.

Here, we consider a minimal model with
a transformation rule possessing both a self-referential structure
and some driving mechanism leading to diversity.
For simplicity, we restrict the initial network to that represented by a 
one-dimensional map, i.e., a one-variable function.
This means that each point \(x'\) is mapped to a point \(f(x')\).
In other words, the function \(f(x)\) represents the connection 
between a point in the inarticulate network \(x\) and another point \(f(x)\).  The function $f(x)$
determines a network of relations among words.  By setting
some initial relations through an initial function,
we study how the network spontaneously grows and generates
articulation from the initial inarticulate network.
As the dynamics of the network, we postulate that the form of the function changes  
through reference to the function itself.  This self-reference is represented
by the application of the function to itself, given by the
term \(f \circ f(x)\). 
Another and equivalent possible interpretation may be made by
regarding the function $f(x)$ as
a filter from the input $x$ to the output $f(x)$.
In a biological system, the filter is changed in time
by some feedback process from its output.  We try to 
capture the nature of the self-feedback process due to the output
and its influence on the filter itself, as the simplest form of
self-reference.
\footnote{
  In the case of a filter, the term $ f_n \circ g(x)$
may be more relevant, with a given external function $g(x)$
representing the nature of the external world (environment).
Indeed, we have carried out some simulations for such a model,
but the class of phenomena and concepts to be presented here
appear in this case, too.  Another extension necessary with the above
interpretation may be the use of sequential dynamics rather than
parallel change for all $x$.  As mentioned in Sec.11, some structures
to be presented are also relevant to this extension.}
Now the evolution of the function is written as follows:

\begin{equation}
f_{n+1}(x) = F(f_n (x), f_n \circ f_n (x)).
\end{equation}
The term \(f_n \circ f_n (x)\) changes the connections 
from \(x \rightarrow f(x)\) to \(x \rightarrow f\circ f(x)\) for all \(x\).
Next, we assume that the change of $f(x)$ vanishes when the
self-reference of a function agrees with the function itself.  In other words, we assume that the function can
`relax' to a fixed point function satisfying $f(x) = f\circ f(x)$.

For example, when one listens to a sound $x'$ through the filter $f_n(x)$ and pronounces $f_n(x')$,
it is referred as $f_n\circ f_n(x')$. 
In this case, the relation $f(x') = f\circ f(x')$ represents 
a self-consistent relation for the imitation of sound. 
If $f(x) = f\circ f(x)$ is satisfied as a whole function, 
the network is articulated to form a consistent input-output table.
\footnote{The function $f_n(x)$ is an abstraction of the network 
and is not directly a proposition satisfying this relation. 
The proposition is given as a combination of articulated intervals \cite{II}.}

As the simplest form of the evolution, we choose the form
\begin{equation}
F(x, y) = (1 - \epsilon) x + \epsilon y,
\end{equation}
which is nothing but the operation of a weighted mean with weight parameter \(\epsilon\).
The evolution equation of the function is now given by
\begin{equation}
f_{n+1}(x) = (1 - \epsilon)f_n (x) + \epsilon f_n \circ f_n (x).
\end{equation}
The time evolution of these functional dynamics is determined by 
giving an initial function $f_0(x)$ and a control parameter $\epsilon$.
In other words, the inarticulate network spontaneously evolves without referring to the outer world.
The outer world is given as an initial function $f_0(x)$.
The complexity of the outer world ($f_0(x)$) 
and the intensity of self-reference $\epsilon$ determines the time evolution.

Although we have introduced the evolution rule of the function $f_n(x)$ as Eq.(3),
this rule is applied within the functional space.
We can study how some points $f_n(x')$ work as rules for other points, 
through the application of the function to itself by (3).
Indeed, the emergence of rules from objects will be demonstrated in a subsequent paper.


Depending on the initial function \(f_0 (x)\), the final state
of the function as $n \rightarrow \infty$ is generally different.
These different $f_n(x)$ represent the variety of manners of articulation as a network of relations.
Here we study the evolution of the function under this iteration, 
varying the initial function \(f_0 (x)\) and the (control) parameter 
\(\epsilon\).  In the present paper, 
we choose a monotonically increasing 
function, 
a logistic map (as a representative of single-humped map), as
the initial function \(f_0 (x)\), while  
some other \(f_0 (x)\)  leading to periodic points will be briefly
discussed in Sec.9.

%
%
%

A model related to with ours, 
including the $f \circ f$ term, was proposed by Deutsch \cite{Deutsch1} \cite{Deutsch2}.
In that model motion in a random potential is studied under the iteration of functions.

\section{Recursive Equation with $\epsilon =1$}

First, we discuss the case  
\(\epsilon = 1\), in which the dynamics of the network are simple.
The equation is written 
\footnote{
    The functional equation with $\epsilon =1$ has some similarity with
the renormalization group (RG) equation for the period-doubling bifurcation \cite{Feigenbaum} when
we choose a quadratic function for $f_0(x)$.
In contrast with the case of the RG equation, however, a scaling transformation is not included in our
model equation.}

\begin{equation}
f_{n+1} (x) = f_n \circ f_n (x).
\end{equation}
This iteration yields snapshots of \(2^n\) steps of the
equation $g_{n+1} (x) = f_0 (g_n (x))$, $g_0(x) = x$.
If \(f_0 (x)\) is a map which generates a chaotic orbit,
equation (4) generates a chaotic sequence.

As a simple introduction, let us consider the `discrete mesh' case, where
the initial function $f_0(x)$ takes only $M$ possible values,
(for example given by $f(i)=j$ $(i, j = 0,\ldots,M-1)$.  
We adopt the integer index $i$ for each element in this
discrete mesh, and we denote by $f(i)$ the functional value at $i$.
In this case, we can consider only
$M$ elements, and the function is nothing but a network from
$[0,1,\ldots, M-1]$ to the same set.
Once a map, represented by a finite mesh, is given, the model gives
the dynamics of the network in which each element connects to
itself or to a different element.
Each element is represented by a site index
$i$, while $f_n(i)$ gives the site index to which $i$ is mapped.

In this equation, the functional map changes only
the connection from the element  \(f_n(i)\)
to the element to which the
mapped element \(f_n(i)\) is mapped, that is \(f_n \circ f_n (i)\).

In this section we discuss only the special case that $M$ elements are arranged cyclically as
$f_0(i) = i+1 \pmod M$ (for $i = 0, \ldots, M-1$).
General properties of the network with a `discrete mesh'
are discussed in the Appendix.

The evolution of the network of $M$ elements is
displayed in Fig.\ref{fig:e0peri} for $M=1,2,\ldots, 7$.
We call the network `elementary' if it does not
disintegrate into parts upon iteration of the function.
A one-element cyclic network (i.e., type-I fixed point) is obviously `elementary'.
A cyclic network with $M=2$ is reduced to two disintegrated fixed points,
and thus is not `elementary'.
A cyclic network with $M=3$ has a period-2 cycle without disintegration
and is an elementary network.

First, note that a network of $2m$ cyclic elements
is reduced to an \(m\)-element network, since the first
iterate of the map leads to connection to the
next nearest neighbor and produces two disintegrated networks of 
\(m\) elements. Repeating this process, cyclic networks of
\(2, 4, 8, \ldots, 2^n\)  elements are finally
reduced  to fixed points.
In the same way, networks of \(6, 12, \ldots, 3 \cdot 2^n\) elements are
reduced to elementary networks of three elements with
period 2.  We thus see that a network consisting of an even number of elements
cannot be `elementary.

Contrastingly, a cyclic network of an odd number ($M$) elements is
not reduced to disintegrated elements. 
Such a network remains a cyclic one as a results 
of the  first iterate. 
Therefore, the network does not disintegrate under the
next iteration, and this argument can be repeated ad infinitum. A network consisting of
\(M\) elements is rearranged and returns to the original position 
after some number \(n\) of iterations (\(n < M\)).
Therefore cyclic networks of odd numbers are elementary.

The period $P(M)$ is plotted for odd $M$ in Fig.\ref{fig:p10000}.
For some values of $M$ the period assumes the maximal possible value
 $M-1$ (e.g., for $M=3,5,11,\ldots$), 
while a sequence of some numbers $M$, satisfies $P(M)=$ $(M-1)/2$, $M/3$, \((M-1)/4\) and so forth, 
as shown in Fig.\ref{fig:p10000}.
The algorithm to determine $P(M)$, as well as its upper bound, is given in the Appendix,
where it is also shown that any network is attracted into a combination of elementary networks.

For functional dynamics with a countably infinite number of mesh points,
the points may not be attracted into an elementary cycle
within a finite number of time steps.  However, the network structure
described here for a finite mesh exists in the case that
element number is countably infinite.  
For a real number $x$, a chaotic orbit can exist, as mentioned at the beginning of this section.
However, if $f_0(x)$ is a map with periodic attractors,
the dynamics are expected to be defined in terms of these of elementary networks with fixed points 
and transient behabiour exhibited in the evolution toward these points.


Note, however, that the structure of an elementary network described here has
no correspondence to the case with $\epsilon <1$.

\section{General Properties}

Now, we consider equation (3) with \(\epsilon \neq 1\): 
\begin{equation}
f_{n+1} (x) = (1 - \epsilon) f_n (x) + \epsilon f_n \circ f_n (x).
\end{equation}

In this paper, we consider only functions whose ranges are subsets of their domains.
Such a function is bounded from above and below by the above mapping of 
\(f(x)\) with \(0 \le \epsilon \le 1\).
By rescaling $x$, we can choose the domain  
of a function in \(0 \le x \le 1\) (which contains its range as a subset).

The evolution described by our model is characterized by 
\(\epsilon\) and the initial function \(f_0 (x)\).
Also, we note that 
two points with the same \(f(x)\) at some time value exhibit identical evolution subsequently,
since our model is completely specified by \(f(x)\).

In the following study, it is useful to introduce the concept of the 
``self-contained section" (SCS).
The SCS is defined as a connected interval $I$ such that
$f(I) \subset I$, no connected interval $J \subset I$
satisfies $f(J) \subset J$, and $f(I+\delta)$ does not
satisfy $f(I+\delta) \subset I+\delta$ for arbitrarily small $\delta$.
The total interval $[0,1]$ may include several SCS.
In each SCS the function $f_n(x')$ is mapped into the SCS.
Thus, $f_{n+1}(x)$ remains in the SCS.

For a given function \(f_n(x)\), 
the domain can be divided into SCS intervals and points outside these intervals.
For an SCS $I$, the evolution of $f_n(x)$ for any $x \in I$ is determined completely
by the evolution of $f_n(x)$ within this interval alone.
Information regarding the
evolution of $f_n(x)$ in an SCS is self-contained. 
The evolution of the remaining parts, on the other hand, is not self-contained
but is affected by $f_n(x)$ in the SCS to which $x$ is mapped.\\

This form of the functional map (5) has, of course, a fixed point solution with
$f(x)=f \circ f(x)$.  Even if this solution may not be satisfied for all
values of $x$, the fixed point condition is often satisfied locally at some
points $x$.  There are two types of such fixed points.
(Note that this does not necessarily mean 
that the function is a fixed function for the whole domain.)

(i) a point $x^I$ satisfying \(f (x^I) = x^I\) (Type-I).

(ii) a point $x^{II}$ satisfying $f(x^{II}) = f \circ f (x^{II})$
(but $f (x^{II}) \neq x^{II}$) (Type-II).



A type-I fixed point $x^I$ is such that \(x^I\) is mapped to \(x^I\). 
A type-II fixed point, given by the condition
\(f (x^{II}) = f \circ f (x^{II})\) but
\(f (x^{II}) \neq x^{II}\) depends on a type-I fixed point:
\(f (x^{II}) = f \circ f (x^{II})\) implies that 
$f(x^{II})$  is a type-I fixed point.
Thus, in reference to Fig.\ref{fig:evolve}, the type-II fixed points are those with the same
heights as the type-I fixed points.

The type-II fixed point $x^{II}$ is mapped to $f_n(x^{II})$, where it remains under subsequent evolutions, 
as $f_n(x^{II})$ is a type-I fixed point.
As $n$ increases the number of points at which $f_n(x)$ intersects the identity function increases.
Thus, most points of \(x\) are expected to converge to a fixed point,
especially within a finite mesh simulation of a functional map.
Still, periodic points (or chaotic points)
also exist. This point is 
investigated in Sec.9 and in a subsequent paper \cite{II}.\\

In the rest of the present section we study the simplest case, i.e.,
the evolution from an initial, continuous monotonically increasing function (Fig.\ref{fig:mono}).
In this case, $f_n(x)$ converges to a step function as $n \rightarrow \infty$.
Note first that if \(f_n (x)\) is continuous and monotonically increasing, 
\(f_{n+1} (x)\) is too.
Further, if \(f_n(x^\prime) > x^\prime \), 
\(f_n \circ f_n (x^\prime) \ge f_n (x^\prime)\) also holds.
Thus \(f_{n+1} (x^\prime) \ge f_n (x^\prime)\).
Since \(f_{n+1} (x)\) conserves the monotonically increasing property,
a given initial function  intersects the 
identity function at some points. 
Let us denote by $x_i$ successive intersection points of an initial function
\(f_0(x)\) with $x_i<x_{i+1}$ ($f_0(x_i)=x_i$).
Thus $f_n(x_i)$ is a type-I fixed point.
Hence, the function \(f_0(x)\) can be decomposed into some sections $[x_i,x_{i+1}]$. 
In these sections, $f_0(x)$ satisfies either $f_0(x) \ge x$ or $f_0(x) \le x$.
In the former case, a slightly smaller interval $I \equiv [x_i + \delta, x_{i+1}]$ 
satisfies $f_0(I) \subset I$, and
we can choose a function $h_n(x)$ so that $h_n(x) = \gamma_0$ as $\gamma_0 (x - x_{i+1}) + x_{i+1} 
\le f_0(x) \le x_{i+1}$ with $0 < \gamma_0 < 1$ for $x \in I$.
If $h_n(x) < f_n(x)$, then $h_{n+1} < f_{n+1}$ holds,
since $h_n(x)$ and $f_n(x)$ are continuous, monotonically increasing functions.
For arbitrary $n$, the relation that $\gamma_n (x - x_{i+1}) + x_{i+1} \le f_n(x) \le x_{i+1}$ is satisfied.
Here, $\gamma_{n+1} = (1-\epsilon)\gamma_n + \epsilon \gamma^2_n$.
Hence, $\lim_{n\rightarrow\infty} \gamma_n = 0$, 
and thus $\lim_{n\rightarrow\infty} f_n(x) = x_{i+1}$ (type-II fixed point) uniformly on $I$.
In the latter case, with $f_0(x) \le x$, 
an interval $I \equiv [x_i, x_{i+1}-\delta]$ satisfies $f_0(I) \subset I$ and vice versa.


Hence, the function $f_n(x)$ as $n \rightarrow \infty$
converges to a step function consisting of fixed points, and 
it is articulated into each interval in which 
\(f(x)\) takes the same value (see Fig.\ref{fig:mono}).
A domain in which $\lim_{n \rightarrow \infty} f_n(x)$ assumes a single
value is given by the connected interval $(x_i, x_{i+1}]$ or $[x_i, x_{i+1})$.
The set of such domains 
is determined once we choose an initial function.

It is clear that the approach to this step function is
independent of the value of \(\epsilon\), which changes only 
the speed of convergence.

The present example demonstrates the simplest evolution of
our functional map.  For a general initial function, it is not easy to
make analytic arguments to understand the qualitative nature of the evolution, and one has to
resort to numerical simulation. For a simulation we have to 
divide the interval \([0.0, 1.0]\) into a finite number of mesh points.
This use of a finite mesh is equivalent to the use of a piecewise constant 
function whose values $f(x)$ are restricted to $i/M (i = 0,1,\ldots, M)$ 
with a large integer $M$ giving the mesh size $1/M$.
As an approximation of the evolution of a smooth initial function $f(x)$, 
use of a finite mesh size may introduce an artificial effect.
In particular, if a function intersects the identity with a large slope $|f'(x)|$,
the corresponding type-I fixed point may be overlooked in the 
finite mesh simulation.  Another effect is seen near a tangent bifurcation
with a type-I fixed point when the slope $f'(x)$ is close to 1.
In this case, instead of a single type-I fixed point, the finite mesh
simulation may have a tendency to produce a chain of type-I fixed points
continuing over some interval.
We treated such problems in a statistical manner,
increasing and decreasing the mesh size (e.g., $\pm 5$) and computed averaged quantities
for such meshes, to determine the mesh size dependence.

\section{Functional Logistic Map}

As a nontrivial class of evolution, we choose a logistic map 
as an initial function.  
Indeed the behavior to be discussed here is observed
for any single-humped function.  We study the logistic map as
a representative of the universality class
of single-humped maps $f_0(x)$.
We choose the initial function

\begin{equation}
f_0 (x) = rx(1-x),
\end{equation}
with \(x \in (0.0, 1.0)\), \(r \in (0.0, 4.0)\).

We have carried out extensive simulations of this model,
by changing  the parameter \(\epsilon\)
and the configuration of the initial network determined by \(r\).
With this type of the initial function,
the function $f_n(x)$ converges to fixed functions
for small \(\epsilon\) and \(r\). 
If \(\epsilon\) and \(r\) are sufficiently large, 
\(f_n (x)\) does not necessarily converge to a fixed function.
Some points $x$ exhibit periodic behavior, while most of them converge to
fixed points.

An example of the time evolutions is shown in Fig.\ref{fig:evolve}.
With the iteration \(f_n \circ f_n (x)\) and the chaotic dynamics
in the logistic map, the function is folded repeatedly.
Within the first few steps, the function forms several mountains and valleys,
with this folding mechanism. On the other hand,
due to the weighted average of \(f_n (x)\) and \(f_n \circ f_n (x)\) composing $f_{n+1}(x)$,
the function is distorted from the case with $\epsilon = 1$. This
leads to relaxation to a fixed-point structure. 
With the time evolution, the number of
type-I and type-II fixed points increases with successive
folding of the function.
With this creation of fixed points,
the folding leads to form many step structures.

Functions to which $f_n(x)$ converges as 
$n \rightarrow \infty$ can be classified into some types.
Typical such functions are presented in Fig.\ref{fig:type},
for different values of \(\epsilon\) and \(r\).
In this plot of \(f_n (x)\), simulations are carried out with the
mesh size \(= 4096\), where the function converges to a fixed point function
at the time step $n$ on the order of 100.
The function consists of flat pieces and sharp steps.
The flat pieces are derived from type-II fixed points. 
In contrast with the case of a monotonically increasing function, there can be several
separated domains of \(x\) with the same value \(f(x)\).

As shown in Fig.\ref{fig:type},
fine step structures appear in (b) (around $x = 0.0, 1,0$) and 
infinitely fine step structures appear in (c) and (d).
There, finer and finer folding structures appear in time with the folding.
The number of fixed points increases with time, 
although the new structures become successively smaller.
(Note, however, in a  finite mesh simulation, the function 
converges to a function with a finite number of steps).

In Fig.\ref{fig:type}(a) and (b), \(f_n (x)\) converges to a fixed step function.
In Fig.\ref{fig:type}(b), \(f_n (x)\) has localized
fine structures in addition to the steps, 
but as seen in the next section, the function finally converges to a step function. 
In Fig.\ref{fig:type}(c) and (d),
\(f_n (x)\) has infinitely small folds.
As \(\epsilon\) or \(r\) increases,
regions with fine structures start to dominate as seen in (c) and (d).
Flat pieces remain for some intervals in $x$ in Fig.\ref{fig:type}(c), while
almost all regions are non-flat in Fig.\ref{fig:type}(d).

According to overall results of the simulations, the functions to which $f_n(x)$ converges can be classified
into four types.
\begin{itemize}
\item
(S): Step Phase (Fig.\ref{fig:type}(a))
 
In this region \(f_n (x)\) has 3 or fewer  
values, and it converges to a step function with a finite (few)
number of discontinuous points. 

\item
(FS): Folded Step Phase (Fig.\ref{fig:type}(b))

For most intervals of $x$,
\(f_n (x)\) assumes the form of a step function, as in the case (S),
but there are few points $x$ (around $x = 0.0, 1.0)$ around which
a fine folding structure exists.

\item
(F): Fractal Phase (Fig.\ref{fig:type}(c))

The function has flat pieces and some areas with infinitely fine 
folding structures.
In these areas, the folding structure folds itself.
The number of type-I fixed points
increases with iteration.

\item
(R): Random Phase (Fig.\ref{fig:type}(d))

Flat pieces in the function vanish and are replaced by infinitely fine folding structure.
\end{itemize}
In the next section, we discuss the origin of the changes
of these phases.

\section{Mechanism of Phase Changes}

Folding plays a central role in the phase change.
In studying this folding structure, let us recall the concept of 
self-contained sections (SCS).
There is a difference in the folding mechanism between $f(x)$ within an SCS and outside of the SCS.
A part of \(f(x)\) within each SCS evolves by self-folding 
(\(f_n \circ f_n(x)\)).  The balance between self-folding and averaging with $f_n(x)$
determines the shape of the function within the section.
Whether the humped function grows into a sharp step structure or 
a flat piece depends on this balance.

On the other hand, the points outside of the SCS  
are eventually mapped to values \(f_n \circ f_n(x)\) in some SCS, where they remain for all subsequent times.
Following hump(s) in an SCS, the function in the
remaining part can also be folded, as shown in 
Fig.\ref{fig:fff}. The hump in the SCS can lead to a successive folding structure
within itself and also in the remaining part, whose $f(x)$ is mapped to the
SCS.  This
SCS plays an  important role in classifying the types of functions.

The limiting function $\lim_{n \rightarrow \infty}f_n(x)$ is plotted as a function of $r$ and $\epsilon$
in Fig.\ref{fig:bif}.
Here, one dot in the figure represents a value of $f_n(x)$ in  a
flat piece or on a peak of the map. This figure looks
similar to the bifurcation diagram of the logistic map.
In fact, this graph is identical to the 
bifurcation diagram of the logistic map for the case\(\epsilon = 1\).
Although the `bifurcation diagram' is distorted
because of the average by the weight \(\epsilon\),
we can detect  some similarity with the original bifurcation diagram.

In the plot, one can find structures corresponding to tangent bifurcations and other
bifurcations.  Also, there is
'bifurcation collapse' corresponding to crisis \cite{Grebogi}.
Here, however, there is one significant difference.  In strong
contrast to the case of crisis, `bifurcation collapse' occurs at 
different parameter values, depending on each bifurcated branch of $f(x)$
(see Fig.\ref{fig:bif}).
Thus, there is a new regime in the `bifurcation diagram' where
one branch collapses and the other remains stable.
  
To see what happens in this new regime, we plot two functions \(f_5 (x)\) 
before and after a collapse of one bifurcation branch.
As is seen in Fig.\ref{fig:dis}(a), when bifurcated branches coexist,
there are two closed SCS.
Here each SCS corresponds to a bifurcated branch.
Any two branches evolve independently,
because points in an SCS, plotted by the dotted squares  
in the figure, remain within the same SCS.
Regions outside the two SCS are eventually mapped to these SCS.
The folding structure in each SCS is folded by itself while
the folding structures outside these SCS are folded by these SCS.

As $r$ or $\epsilon$ is increased, the bifurcation collapses.
As shown in Fig \ref{fig:dis}(b), one of the the SCS ( around $.3<x<.7$)
collapses while the other remains.
Some parts of the function $f(x)$ at the collapsed SCS are mapped to
the SCS around $x \approx 0.8$.  
This is displayed in (b), where the function in the lower section
`invades' the  upper area.
This parameter region is nothing but the region
where only one bifurcated branch is collapsed.

With further increase of  $r$ or $\epsilon$, the SCS around $x\approx 0.8$
also collapses and
they form a single SCS.  This parameter regime
corresponds to the region in which both the two branches are
collapsed.

The change of the four phases can be understood in terms of  the folding and SCS as follows.

\begin{itemize}

\item
(S) \& (FS): As is seen in Fig.\ref{fig:fold} (a), the folding structure in an SCS
continues to fold the rest of the map
until it converges as  \(n \rightarrow \infty \). If the folding 
structure converges to a step function,
the number of type-I fixed points
no longer increases  for large \(n\).

\begin{itemize}
\item
(S): In the step phase, the folding structure in the each SCS converges to flat pieces.
The points outside the SCS are not folded by the SCS.

This phase is seen in 
the parameter region between the tangent bifurcation and the second bifurcation.

\item
(FS): In folded step phase, the folding mechanism stops at an SCS.
Until the folding structure converges, 
a few points, mapped to an SCS, continue to
form foldings, due to step structures in the SCS.
For most intervals of $x$,
\(f_n (x)\) is a step function, as in the region(S),
while a few intervals exhibit fine folded structure (with a finite number of folds).

This phase is observed in the parameter region between the second bifurcation
and the first bifurcation collapse.
\end{itemize}

\item
(F) \& (R): The folding structure in an SCS folds itself 
when $r$ or $\epsilon$ is large.
Here, as $n$ increases, the function $f_n(x)$ intersects the identity function an increasing number of times.
Thus, the number of type-I fixed points
increases with iteration in the SCS.

\begin{itemize}
\item
(F): In some regions of
the remaining part, the folding does not continue. Here flat pieces
are formed. 
This process leads to a fractal phase.

This phase exists in the parameter region between the first 
and the last bifurcation collapses.

\item
(F)$\rightarrow$(R): With the increase of \(r\),  the number of bifurcation branches,
as well as the number of collapsed bifurcations, increases.
Infinite folding structure starts to cover the entire domain.

\item
(R): In the random phase, all bifurcation branches are collapsed, 
and no SCS smaller than the total interval exists anymore.
The entire domain forms a single SCS with self-folding structure.
Hence no flat pieces remain anymore.

This phase corresponds to the region from the last bifurcation collapse.

\end{itemize}
\end{itemize}

\section{Characterization of Phases}

In the present section  we quantitatively characterize the four phases,
considering the folding effects discussed in Sec.7.  As statistical characteristics, we
compute the number of discontinuous points \(D\) and  
the Euclidean length \(L\) of the fixed point function to which $f_n(x)$ converges after
transient time steps.  We then study 
the length of the transient
time \(T\) before the function converges to an attractor.
Discontinuous points are the edges of the flat pieces of the map,
i.e., the points  where $f_{\infty}(x+1/M) \neq f_{\infty}(x/M)$, while
the length is given by 
$\sum_j \sqrt{(1/M)^2+(f_{\infty}(x_j+1/M)-f_{\infty}(x_j))^2}$.

For each simulation, we choose
the mesh number $M$, the function is iterated until it converges,
and the numbers \(D\) and \(L\) are determined.
The dependence on $M$ is studied as the number of mesh points is
doubled.\footnote{
To avoid complicated dependence on $M$ due to 
finite size effects, the number plotted for each $M$ is the value averaged over the results with 
$M-2,M-1,M,M+1,M+2$ mesh points, as discussed above.}

The log plots of these quantities are displayed as functions of the log of
the number of mesh points $M$ in Fig \ref{fig:quan}.
According to the behavior of these quantities and the previous discussion on the folding structure,
the four phases in Sec.7 are characterized as follows
(See Fig.\ref{fig:phase}).

\begin{itemize}

\item
(S):  Step (Fig.\ref{fig:type}(a))

\(D = const\), \(L = const\) as $M$ increases.

Since the folding structure does not fold the part outside the SCS,
$D$ and $L$ do not change as the number of mesh points increases.

\item
(FS): Folded step (Fig.\ref{fig:type}(b))

\(D \propto M^\alpha\), \(L \propto M^{\alpha^\prime}\),
(\(\alpha\), \(\alpha^\prime < 1\)) for $M < M_{max}$,
while they approach constants for larger $M$.

The function 
\(f_n (x)\) undergoes a stretching and folding process 
under the iteration, until the folding structure converges.
This folding process leads to a successively finer structure, and
brings about more discontinuous points with the increase of mesh points.
The existence of $M_{max}$ such that for $M > M_{max}$, $D = const$ and $L = const$, is expected, 
because the folding structure converges.
The folded region existing outside the SCS intervals is localized in a narrow interval of $x$, 
and the number of mesh points ($M_{max}$) necessary to observe the convergence is huge.
Thus in Fig.\ref{fig:quan}(a)(b), $D$ and $L$ increase with the number of mesh points, 
but we believe that the increase stops with a further increase of mesh points.
In this region, only a finite number of such
folding structures exists, and these regions are separated from
other fixed points with a step function.

\item
(F): Fractal (Fig.\ref{fig:type}(c))
\footnote{
     A linear functional equation leading to a fractal function is
discussed in terms of  Weierstrass and Takagi functions (see \cite{Ymaguchi-Hata}), while
a nonlinear functional equation (with the term $f^2(x)$ rather than $f \circ f(x)$)
has been discussed in \cite{KK} \cite{TK} in relation with a fractal torus.}

\(D \propto M^1\), \(L \propto M^\beta\), (\(\beta < 1\)).

The function has flat pieces and folded areas, where the folding structure folds itself in each SCS.
The number of type-I fixed points (thus possible values of $f_n(x)$)
increases with iteration.  Because of the folding of folding structure itself, 
there are some discontinuous points in any neighborhood of a point $x$ in an
SCS with the bifurcation collapse.  Thus, \(D \propto M^1\) follows. 

\item
(R): Random (Fig.\ref{fig:type}(d))

\(D \propto M^1\), \(L \propto M^1\).

Flat pieces in the function vanish, and they are replaced by
folded regions.
In this area there are discontinuous points 
around any infinitesimal neighborhood of $x$. Now, the statistical
behavior of $D$ and $L$ is identical to that of a function with random values
at each mesh point.

\end{itemize}

Of course, the number of type-I fixed points (possible values of $f(x)$)
is a basic quantities.  This number does not increase with the mesh number $M$
for (S) and (FS) phases, while it increases with $M$ for (F) and (R) phases.  Roughly speaking,
the increase is proportional to $M$, although there is a large variation around this,
possibly due to some number theoretic complication.

In addition to the characteristics of limiting functions $\lim_{n \rightarrow\infty}f_n(x)$, 
we also study the transient process.
In Fig.\ref{fig:quan}(c), the length of the transient $(T)$ 
before the function reaches a fixed point is plotted.
The transient length stays very small in (S) and (FS) phases.
On the other hand, the length increases in proportion to \(M^{\delta}\),
in (F) and (R) phases.  The exponent $\delta$ increases with $r$
in the (F) phase, and it is approximately 1/2 in the (R) phase. 

This divergence of the transient length
implies that the function does not converge to a fixed point function
in the limit of infinite precision. The process in which a finer 
folding structure is formed continues forever.  Note, however, that our
numerical results with a finite mesh can capture the behavior of the
function. Even though the function does not converge in the
infinite precision limit, the dynamical change at a fine scale does not
affect the larger structure.  As time increases, the change 
in the value of $f_n(x)$ becomes smaller and smaller.
Hence, the classification of $f_n(x)$ obtained with a finite-mesh simulation 
is valid.

Note that the phase diagram (in Fig.\ref{fig:phase}) plotted with the above characteristics 
agrees with that obtained earlier using the bifurcation collapse and
folding characteristics (see Table 1 for a summary).
The phase transition is also characterized by other `order parameters'.
In Fig.\ref{fig:states}, the quantity $\int dx f_{\infty} (x)$
and the number of type-I fixed points (the number of possible states)
are plotted.  From these figures, we can see the bifurcation
of the phases as the parameter value is changed.

The value $\int dx f_{\infty} (x)$ is governed by a dominant step structure for small \(r\).
At the first and second bifurcation points, there are cusps in the
integral.  
After the first bifurcation collapse, the integral does not change
smoothly as a function of the parameter $r$. Rather, it begins to exhibit
sensitive dependence on $r$.  At the last bifurcation
leading to the (R) phase, there is a large jump in the integral, due to the
collapse of the SCS.

Such bifurcation structure is also seen in the
change of the number of type-I fixed points,
 plotted in Fig.\ref{fig:states}(b) (for $M = 4096$).
The number remains small (3 or fewer) up to the second bifurcation.
At the second bifurcation
leading to (FS), it starts to increase slightly.
After the first bifurcation collapse, 
the number jumps to a much larger value.

In the present and last two sections we have discussed the
function dynamics for the case $\epsilon \neq 1$, starting from a single-humped map.  
For \(\epsilon = 1\), our dynamics is nothing but normal iteration of the logistic map.
In this case, `bifurcation collapses' occur at the same parameter.  
The function \(f_n (x)\) exhibits fixed, periodic and chaotic behavior, 
as in the bifurcation of the logistic map. (In a finite-mesh computation,
the number of type-I fixed points is much larger than
the case \(\epsilon < 1\), while periodic points are also frequently observed there.)

\section{Periodic Attractor}

To this point, we have focused on the fixed point solutions
in the case $\epsilon \neq 1$.  Although they are not so common,
periodic points of  $f_n(x)$ are also observed in the case $f_0(x) = rx(1-x)$.  In fact, 
it is often the case that a function with multiple humps evolves into a function possessing periodic points.
Even in such case, the function $f(x)$ is a fixed point for most points, 
and only at a few points $x'$ $f_n(x')$ changes periodically.  
Here we study how such periodic points are constructed in the case that
they depend on a finite number of type-II fixed points. \footnotemark
\footnotetext{
When there are an infinite number of type-II fixed points, the function dynamics
can generate  some dynamic rule, as exists in the systems mentioned in the context of the 
third and fourth problems in Sec.1.  
These dynamics have more variety, including
chaos, or `meta-chaos', as will be reported in a subsequent paper\cite{II}. 
`Meta-chaos' consists of dynamics in which the evolution rule itself changes chaotically.
These dynamics have stronger orbital instability than chaos, in the sense $exp(n^2)$ or faster with time $n$.}

The mechanism allowing for a periodic cycle for \(\epsilon \neq 1\) is different 
from that for the case with \(\epsilon = 1\), discussed in Sec.4.
Here, we focus on the case \(\epsilon \neq 1\).
Periodic attractors for a particular $x'$ with an arbitrary period are constructed as follows.
We note that as far as we have examined extensively, a periodic point moves 
successively on type-II fixed points.

For example, let us design a period-2 solution with $f_{2n}(x')$ ($f_{even}(x')$),
$f_{2n+1}(x')$ ($f_{odd}(x')$) and $f_{m+2}(x')=f_{m}(x')$.
Our purpose is to arrange the fixed points so that $f_n(x')$ is mapped to two different fixed points by each step.
If $f_n(x')$ is mapped to a type-I fixed point, $f_{n}(x')$ becomes a type-II fixed point.
Here, we treat such a case that $x'$ is mapped to a type-II fixed point.
In this case, to have period-2 solution of $f_n(x')$, 
2 type-I fixed points and 2 type-II fixed points must be chosen.
We denote type-I fixed points as $a_1$ ($f(a_1) = a_1$) and $a_2$ ($f(a_2) = a_2$)
and the correspondent type-II fixed points as $a_1'$ ($f(a_1') = a_1$) and $a_2'$ ($f(a_2') = a_2$).
Now we consider the  situation shown in Fig.\ref{fig:peri}(a), where
it is assumed \(f_{even}(x^{\prime}) = a_1'\) and \(f_{odd}(x^{\prime}) = a_2'\). 
The condition for period-2 is given by

\begin{equation}
  \left\{
    \begin{array}{lclcl}
       f_{odd} (x^\prime)    &=& (1 - \epsilon) f_{even} (x^\prime) &+& \epsilon f \circ f_{even} (x^\prime)\\
       f_{even} (x^\prime)   &=& (1 - \epsilon) f_{odd} (x^\prime)  &+& \epsilon f \circ f_{odd} (x^\prime)
    \end{array}
  \right.
\end{equation}
which is written as
\begin{equation}
  \left\{
    \begin{array}{r}
       a_2' = (1 - \epsilon) a_1' + \epsilon a_1\\
       a_1' = (1 - \epsilon) a_2' + \epsilon a_2
    \end{array}
  \right.
\end{equation}
From this equation, \(a_1, a_2\) are determined by \(a_1^\prime, a_2^\prime\) (or vice versa) as 

\begin{equation}
  \left(
    \begin{array}{c}
       a_1'\\
       a_2'
    \end{array}
  \right)
       = \frac{\epsilon}{1 - (1 - \epsilon)^2}
  \left(
    \begin{array}{cc}
       (1 - \epsilon) & 1\\
       1 & (1 - \epsilon)
    \end{array}
  \right)
  \left(
    \begin{array}{c}
       a_1\\
       a_2
    \end{array}
  \right)
\end{equation}
If this condition is satisfied for \(a_1\),\(a_2\),\(a_1'\) and \(a_2'\),
$f_n(x') = f_{n+2}(x')$.

A function of an arbitrary period is constructed in the same way as
one of period 2.  To construct a solution of  period \(N\), we introduce
$N$ type-I fixed points denoted by $a_i$ $(i = 1,2, \ldots, N)$ and 
$N$ corresponding type-II fixed points denoted by
\(a_i' (i = 1,2,\ldots, N)\) with \(f_n(a_i') = a_i (i = 1,2,\ldots,N)\).
For \(f_n (x')\) to change the values $a_i'$ $(i = 1, 2, \cdots, N)$ cyclically,
the fixed points have to satisfy the condition

\begin{equation}
  \left(
    \begin{array}{c}
       a_1^\prime\\
       a_2^\prime\\
       \vdots\\
       a_N^\prime
    \end{array}
  \right)
       = \frac{\epsilon}{1 - (1 - \epsilon)^N}
  \left(
    \begin{array}{ccccc}
       (1 - \epsilon)^{N-1} & (1 - \epsilon)^{N-2} & (1 - \epsilon)^{N-3} & \ldots & 1\\
       1                    & (1 - \epsilon)^{N-1} & (1 - \epsilon)^{N-2} & \ldots & (1 - \epsilon)\\
       \vdots               & \vdots               & \vdots               & \ddots & \vdots\\
       (1 - \epsilon)^{N-2} & (1 - \epsilon)^{N-3} & (1 - \epsilon)^{N-4} & \ldots &(1 - \epsilon)^{N-1}
    \end{array}
  \right)
  \left(
    \begin{array}{c}
       a_1\\
       a_2\\
       \vdots\\
       a_N\\
    \end{array}
  \right)
\end{equation}.\\
which is obtained in the same way as in the period-2 case.

If an initial function is continuous and has the type-I and II fixed points $a_i$ and $a_i'$ for $i=1,\ldots,N$,
there are some points $f_n(x')$ which are period $N$.
An example of the shape of such a $f_0(x)$ is 
displayed in Fig.\ref{fig:peri}(c) for the case $N=3$. 
The map has period-3 points plotted with small black points.
If all $a_j$ except one are $0$, 
the function of period $N$ has \(N\) hills and \(N\) valleys.
Indeed, an \(N\)-humped initial function has the potential
to possess period-$N$ points.\footnotemark

\footnotetext{Even by starting from a single-humped initial function,
a function with two humps can be formed at the next step if $\epsilon$ is
not small.  Hence, an initial single-humped function also has 
potential to form periodic points, and in fact we have observed a few such cases 
in simulations.}

We have carried out simulations starting from an initial function with many
humps.  As expected, the limiting function consists of
type-I and type-II fixed points, that form a step, folded step, fractal,
and random phases, depending on $\epsilon$ and the height of the humps.
Within these structures, periodic points are embedded. For several
values of $x'$, the function $f_n(x')$ falls on the same periodic cycle
mapping the same type-II fixed points. 
In Fig.\ref{fig:peri}(d), the return map starting from Fig.\ref{fig:peri}(c) 
for all $(f_n(x'), f_{n+1}(x'))$ is displayed ($n = 1000 - 1010$).
This return map was produced given by a computer simulation with $M= 8000$.
All $f_n(x')$ fall onto a fixed point or onto a period-3 attractor, and $f_n(x)$ is a period-3 function as a whole.

The values of the period-3 function $f_n(x')$ for several points $x'$ often oscillate synchronously
with the same phase.  Before $f_n(x)$ falls onto a periodic point,
the function often takes the same $f_n(x)$ value for several values of $x$.
Later, the function $f_n(x)$ is mapped to type-II fixed points, and
starts to form a cycle.
In this case, all $f_n(x)$ for these $x'$ values oscillate synchronously.

Hierarchical organization of periodic points is also possible.
Noting that for a $k$-periodic point $f_n(x')$ takes the same value every $k$ steps,
we can construct a new periodic point by utilizing other 
periodic points.
First, we select one period-$k$ point.
If another point $f_n(x'')$ is mapped to this point after $k$ steps,
the $k$-periodic  $f_n(x')$ acts as a fixed point for $f_n(x'')$,
and a consistent hierarchical equation can be constructed.
For this, we have to prepare period-$k$ points $x_1',\cdots,x'_k$
that are used to make a new period-$k$ point, where
each $k$-periodic point acts as a fixed point for $x''$ per $k$ steps.
Thus, the required number of $k$-periodic points to make a new one is $k$.
If each period-$k$ point acts as a fixed point for $f_n(x'')$, 
$f_n(x'')$ will also have  period $k$.
Thus, we can obtain a hierarchical periodic point $f_n(x'')$ 
depending on other periodic points.\footnotemark

\footnotetext{Since periodic points discussed earlier are mapped
to type-II fixed points, they may be regarded as type-III.  In this hierarchy,
the periodic points that are constructed here may be regarded as type-IV.
Such hierarchy will be discussed in a subsequent paper in detail.}

For example, we select two period-2 points 
(see schematic configuration Fig.\ref{fig:peri}(b)).
Two period-2 points, $f_n(x_1')$ and $f_n(x_2')$ are mapped to type-II
fixed points
as $f_{even}(x_1') = a_{11}'$, $f_{odd}(x_1') = a_{12}'$, and
$f_{even}(x_2') = a_{21}'$, $f_{odd}(x_2') = a_{22}'$, respectively.
If $f_n(x'') = x_1'$ for even $n$ and $f_n(x'') = x_2'$ for odd $n$,
$f_n(x'')$ is mapped $a_{11}$ and $a_{22}$  alternatively.  This condition is
given by
\begin{equation}
\left\{
\begin{array}{lcl}
f_{odd}(x'')   & = & (1 - \epsilon) f_{even}(x'')   + \epsilon a_{11}'\\
f_{even}(x'') & = & (1 - \epsilon) f_{odd}(x'') + \epsilon a_{22}'
\end{array}
\right.
\end{equation}
The above equation has the same form as equation (9).
If $a_{11}, \ldots, a_{22}$ and $a_{11}', \ldots, a_{22}'$ 
support two period-2 points and 
$x_1', x_2', a_{11}'$ and $a_{22}'$
satisfy the above condition,
$f_n(x'')$ evolves by period-2.

In the same way period-$k$ points which depend on $k$ period-$k$ points can be
constructed.  Also, hierarchical construction for the next level
periodic function can be carried out in the same manner.

In the fractal and random phases,
the evolution leads to infinite type-I fixed points.
In the present paper, however, we have discussed only the 
construction of periodic solutions using a finite number of
type-I and type-II fixed points.  If there is a continuous interval consisting entirely of type-II or
type-I fixed points,  the evolution of the function itself starts to be  governed by some
mapping generated by this interval of fixed points. 
Then, quasi-periodic, chaotic, and `meta-chaotic' evolutions of function are possible, as will
be discussed in a subsequent paper \cite{II}.

\section{Summary}

In the present paper, we have introduced a simple functional map to consider
a dynamical system in which rules and variables (or objects)
are not distinguished at the initial time.
A simple (possibly the simplest) universal model was introduced
to study such a situation, as a map describing the dynamics of a function.

As a first step, we studied the case in which the dynamics are given simply 
by $f \circ f(x)$, which can be thought of as defining the way the network is reconnected.
Its elementary cyclic structures were revealed.  In Sec.4, we illuminated
some of the basic structures of our functional map.  
We found that the type-I and type-II
fixed points provide the elementary core structure.  
A type-I fixed point
is mapped to itself under $f(x)$ ($x^I=f(x^I)$) and forms a basis of symbolization
in the abstract language space ($x$).  
A type-II fixed point $x^{II}$
is mapped to a type-I fixed point under $f(x)$ ($x^I=f(x^{II})$.
The concept of a self-contained section (SCS) was also introduced as a region in which the functional
dynamics remain confined within the the region in question. 

The articulation process studied here is a process in which intervals of $x$
are classified according to how $f_n(x)$ converges to divided intervals
consisting of type-II fixed points,
while each type-I fixed point corresponds to a symbol for each articulated object.
The function as a filter articulates the continuous world $x$
into a set of segments on each of which $f_n(x)$ assumes a distinct constant value.

Starting from a monotonic function, a piecewise constant solution is
reached as a fixed function.  Each step consists of
a continuous set of type-II fixed points
mapped to the same type-I fixed point.
This step function is the simplest example for
the articulation process. In Secs.5-8, we studied the evolution from a
single humped map, where the folding of the function to itself can lead to
many type-I fixed points.   Depending on the degree of folding,  
the limiting forms of $f_n(x)$, $\lim_{n \rightarrow \infty f_n(x)}$ 
are classified as step (S), folded step (FD), fractal (F) and random (R) phases.
These phases are characterized by the mesh number dependence of the number of type-I
fixed points and discontinuous points and of the length of the function defined in Sec.8.
(see Table 1).

In the step and folded step phases, as in the case for a monotonic function,
our functional dynamics lead to a partition of $x$ into a
continuous interval \([x_i, x_{i+1}]\) in which \(f(x)\) assumes a constant value ($ const_i$).
Rigid, fixed articulation structure is formed there.
The difference in the degree of folding distinguishes the (FS) and (S) phases.
In the fractal and random phases, the function successively forms smaller and smaller
articulation structures (Secs.6, 7).  In the fractal phase, successive
foldings are restricted within SCS and $f_n(x)$ outside the SCS is folded by the SCS.
In the random phase, the whole interval folds itself and forms
successively smaller structures.  
In these two phases, finer structures are formed successively,
and a fixed point function is not reached in the
limit of an infinite number of mesh points. 
In Sec.9, we constructed periodic functions using a finite number of fixed points.
Hierarchical periodic points were also constructed by assuming new periodic points $x''$
mapped to a higher-level periodic point $x'$ ($f(x'') = x'$).

\section{Discussion}

Note that the type-I and type-II fixed points provide a basis to have the five
requisites discussed in Sec.1.  These fixed points give
a core structure to the network, obtained through the
iteration process.  One might say, in some sense, that
with the emergence  of type-I and type-II fixed points,
code and encoding become separated.  Such separation is not limited
to these two types of fixed points. The distinction between SCS
and the points outside these provide a separation of self-referenced units and 
the structure mapped to them.
Summing up, the evolution of our dynamics can capture  

\begin{itemize}
\item core structure (prototypes) as fixed points

\item folding structure (categories) in SCS, 
which leads to the articulation 
of the network and alters the folding structure itself 

\item the points outside SCS intervals that are mapped to some SCS (entailed categorization by categories)

\end{itemize}

Note that the above described structure of our model
corresponds well with the separation of cognition and notion,
in language.  
Out of an inarticulate and time-variant network, some 
invariant structures are separated as a rigid structure through iterations.
This rigid structure, at the lowest level, is given by fixed points, while a set of SCS provides such
rigid structure at a higher level.
The configuration of fixed points can provide a base for periodic motion of other points, while
SCS have the role of controlling the remaining, vague part.
With the rigid parts, some structures are articulated. This rigid part provides a basis to
describe the web of relationships or circulation of signs.

When an initial function (a single-humped map) is given, the SCS for the function is the whole domain.
Through iterations, time-invariant parts and time-variant parts can be separated.
The dynamics of variant part (that outside SCS intervals) 
is governed by that of the function within the invariant (SCS) part,
while the invariant part also is mapped to variant parts before it forms the SCS.
During the transient process to form the invariant part,
the dynamics of the invariant part may depend on variant parts also.
In addition to the above roles of the variant part, it can create some relationships between elements within it.

The function $f(x)$ for points $x$ outside the SCS intervals 
can form some relationships through the folding mechanism before the function is mapped into SCS.
For example, synchronization $f_n(x')= f_n(x'')$ can be reached during the transient step
before the function is mapped to a rigid part. 
Although the rigid invariant part governs the dynamics of $f_n(x)$ later, this synchronization relationship,
which is invariant later, is determined only by the dynamics within the variant part.
We may say that the invariant part is the basis to describe the circulation of signs 
(the rigid part being a `stable element' in the sea of relations), 
and that the dynamics of the variant part is determined by the invariant part.
After some iterations, the synchronization is dealt as `social' redundancy by the invariant part.

In fact, within this context,
the term `prototype' was introduced in cognitive linguistics 
\cite{Lakoff}.  For example, language is constrained
by the structure of the human body.
The linguistic structure (prototype and category) suitable for the human body 
(which enables one to iterate language as symbols)
is the foundation for the speech act.
The basic structure of the human body is common for all humankind, and for this reason we have 
common linguistic structure.
This common structure is rigid (invariant) with respect to the iteration, just like the fixed point of our model.
This rigidness provides the possibility of speech act that is common in a society.
Of course, if the entire articulation is common for all individuals, there is no novelty.
The prototype gives only a foundation for language and the category can be articulated in various ways.

Although the speech act is restricted by the condition of the body,
there is some redundancy in the language network with regard to describing something.
In our system, such redundancy is seen in the points lying outside the SCS intervals, where
some `synchronization' is generally observed,
driven identically by a prototype structure in SCS.  Such `synchronization' is
attained through iteration of the functional mapping within the region outside the SCS intervals,
before $f(x)$ is mapped to $x$ in an SCS.

With the change of the bifurcation parameter in our model,
a section is no longer an SCS when a function value in the section starts to be mapped
to outside of the section.  Then, a larger network structure with mutual reference is
formed, as is seen in the collapse of bifurcation in our model.
Such collapse of prototype structure is also a concern of cognitive linguistics.
 
In the fractal and random phases of our model,
successively smaller core structures are formed through the folding process.  Novel 
core structures can be formed ceaselessly in principle.
In these phases, the network exhibits a variety of articulation structures, which have
sensitive dependence on the initial network.  This diversity in the network is
a consequence of our model, where, in contrast to typical  artificial intelligence
studies, rules and objects are not separated in the beginning, and
a table between the two is not given in advance.

It should be noted that the above described structure of prototype and category,
as well as the capacity for novelty  and diversity, are a consequence of
a dynamical system with a self-reference term and initial folding structure.
Such structure is universally observed as long as our dynamics includes
$f \circ f$  and some folding structure, given for example,  by a humped mapping.
In this sense, we may hope that the present study gives a first step
to understand dynamic separation of prototype and category in language,
although the model may be abstract and metaphorical at the present stage.

As mentioned in Sec.1, such separation is not limited to language, but is
often seen in biology, for example, in the separation of function 
between DNA and protein.  Since self-replication processes in life require
a self-referential structure.\footnote{For example, the action of DNA is applied to itself
in replication.}, the present study may give some insight into biological
organization \cite{Rosen}

With folding structure, a function does not always reach a fixed point
for all $x$ values.  In Sec.9, we explicitly demonstrated 
the existence of periodic motion of a function value
by choosing an initial function suitably.
The  value is mapped to several type-II fixed points periodically.
The connection defined by $x \rightarrow f(x)$ dynamically moves over
articulated type-II fixed points 
that are mapped to a `core symbol' (type-I fixed point).
Hence, a dynamic syntax is formed in a hierarchy of periodic points.
In Sec.9, periodic points mapped to such periodic points are also constructed.
Thus, rules over rules can be formed in our model. 
As mentioned in Sec.1,
organization of a meta-rule is  important in biological problems, including
development, cognition, and language.  
In a subsequent paper we will report on how a rule of a one-dimensional map
is formed within our functional dynamics, which allows for
periodic, quasi-periodic, chaotic dynamics.  There it will be shown explicitly
that a rule to change a one-dimensional map itself can be embedded in our
functional equation, which allows for `meta-chaotic' dynamics,
where the rule itself changes chaotically, and the orbital instability is stronger than exponential in time.
With these dynamical structures, 
modalities of rule, meta-rule, meta-meta rule, $\cdots$ will be formed successively.

In this article, we have studied the case with  only one function \(f_n(x)\). In other words,
there is only one self-feedback process for one agent. 
This use of a single function
is, of course, not sufficient to discuss `social' aspects
of language.  For this, functional dynamics with multiple functions $f^i_n(x)$ 
(with the term
$f^i_n \circ f^j_n(x)$) should be considered, to see the articulation and rule-generation
process in a society of agents \cite{III}.  Note that the 
results in the present
paper are for a special case when the functions 
agree ($f^i_n(x)=f^j_n(x)$) through iterations.  Indeed, we have often 
observed such agreement in some preliminary simulations of
the multiple functional dynamics case, where the present argument is 
valid.

Of course, some other extensions should be considered in the future,
including the use of a space of higher-dimension than the one-dimensional space used here,
`sequential dynamics' instead of `parallel dynamics' applied to $x$, the 
addition of noise to smooth $f(x)$, and so forth.  Indeed, 
some preliminary studies suggest that the basic structures 
presented in this paper are valid for these extensions.

Finally, it should be mentioned that self-reference structure is mathematically
studied as domain theory \cite{domain}\cite{Varela}\cite{Rosen}, where consistent and non-trivial sets including 
$f \circ f$ are constructed.  Although there may be some relationship
between our fixed-point functions and domain theory
(and the establishment of such a relationship will be an important future study),
the dynamics in the present approach are missing in domain theory. 
A bridge between dynamical systems theory and domain theory may be
required to construct a mathematical foundation of our functional dynamics, in addition to
the construction of a suitable functional space to support our function.

\appendix
\section{Appendix}

In this appendix we investigate the equation (4) with a `discrete mesh' (see Sec.4).
A simple network which consists of elements arranged cyclically was introduced in Sec.4.
In general, an initial network is given as $f_0(i) = j (i, j = 0, \ldots, M-1)$.
In this equation, the functional map changes only
the connection from the element  \(f(i)\)
to the element to which the
mapped element \(f(i)\) is mapped, that is \(f_n \circ f_n (i)\).
Here we study the dynamics of such a connection changing among a finite number of elements.

A fixed point is classified as either type-I or II, as mentioned in Sec.3.
The value \(f(i)\) for these fixed points remains unchanged through the iteration.

When an initial function (i.e., the connection in the graph) is given, 
the graph is separated into two parts (See Fig.\ref{fig:e01}):

(i) a cyclic network which consists of \(M\)-elements 
(If \(M = 1\), it is type-I fixed point.)

(ii) a line defined as an \(N\)-length line 
which does not belong to (i)
but connects to either (i) or another (ii) line. 

The initial network is represented by a combination of these restricted graphs 
that are drawn in one stroke, since
the number of elements is finite and each element \(i\) has 
one value \(f(i)\).  The allowed graphs are only cyclic networks and lines.
A cyclic network is a set of elements in which a bijection exists.
In other words, \(f_0(i)\) (\(i = 1, 2, \ldots, M\)) takes a different value for each $i$ in the set,
and there is one element $i$ in the set
for each $j$ that satisfies \({f_0}^n (i) = j\) (\(n = 1, 2, \ldots, M\)).
Hence,  the elements of the set are arranged in a cyclic manner.
Each element \(i\) that belongs to this network satisfies
\({f_0}^M (i) = i\), where  \(f^n (x)\) is the \(n\)th
iterate of the function \(f\).
A line consists of \(N\) elements, while 
each element \(i\) is mapped to \(f_0 (i) = i+1\) for \(i = 1,2,\ldots, N\) as displayed in Fig.14.
The last element $N$ is, by $f_0(N)$, mapped to another line or a cyclic network.

If a given initial network possesses some lines, as is easily understood,
the network has at least one line that is mapped to a cyclic network (including a type-I fixed point).
With the time evolution, lines are attracted to the structure to which element $N$ mapped.
Finally, each element which belongs to this line is mapped to an element that
belongs to a cyclic network (i.e., $f^n(i)$  belongs to a cyclic network).
At this stage, elements on the same line evolve in the same way as the
elements in the limiting form of the cyclic network.

Hence, to discuss the behavior of the limiting form of the network (i.e., $f^n(i)$ with large $n$),
we need to consider only the evolution of the \(M\)-element cyclic network.
Under the iteration of the function,                  
the network is either reduced to disintegrated parts or remains
a cyclic network.  We call the network `elementary' if it is not 
disintegrated into parts under the iteration of the function.
A one-element cyclic network (i.e., type-I fixed point) is obviously `elementary'.

\begin{figure}
\noindent
\epsfig{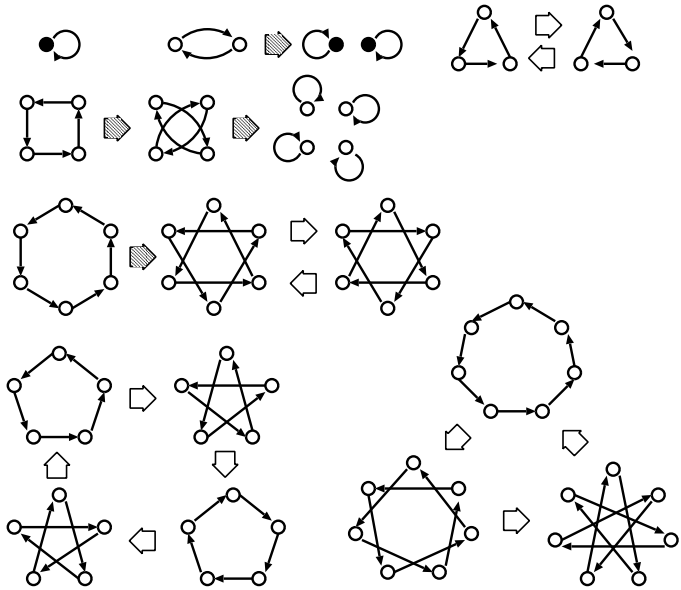}

\caption{
Time evolution of cyclic networks with
\(M = 1, 2, \ldots, 7\). The network with
\(M = 3\) has period 2, that with \(M = 5\) has period 4,
 and that with $M = 7$ has period 3.
The network with \(M = 4\) is reduced to 4 disintegrated fixed points and 
that with \(M = 6\) is reduced to two disintegrated \(M = 3\) networks.
The algorithm to decide the period (see the Appendix) for \(M = 3\) is given by 
\(a_0 = 1, a_1 = 2, a_2 = 1\), and the period is thus 2.
For \(M = 5\), the algorithm proceeds as
\(a_0 = 1, a_1 = 2, a_2 = 4, a_3 = 3, a_4 = 1\), and the period is 4.
}
\label{fig:e0peri}
\end{figure}

\begin{figure}
\noindent
\epsfig{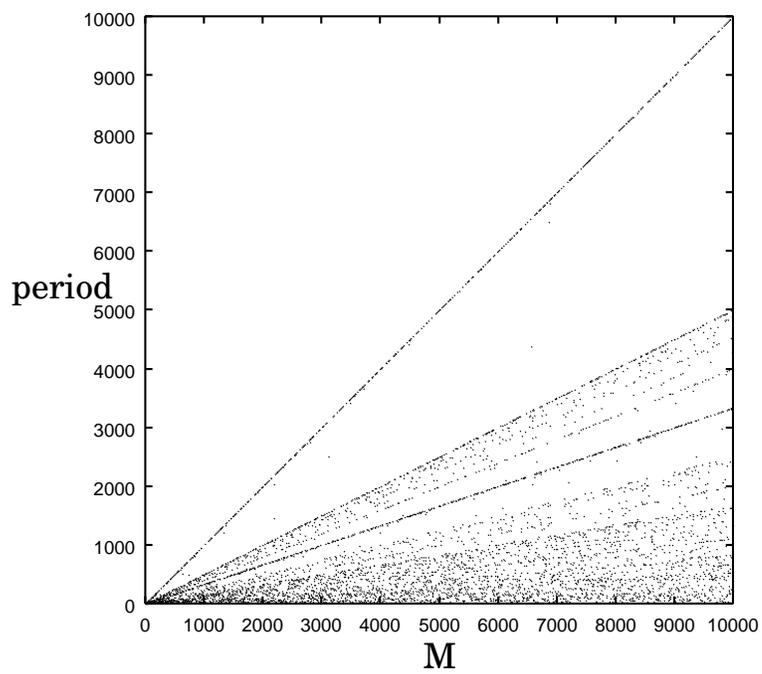}
\caption{The period $P(M)$ plotted for elementary networks  for
odd $n$.
}
\label{fig:p10000}
\end{figure}

\begin{figure}
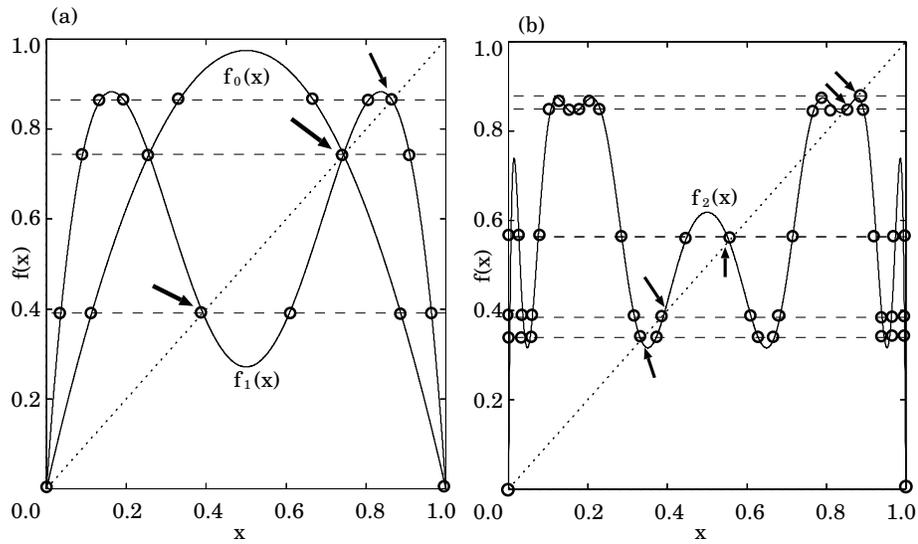

\noindent
\epsfig{file=t01.eps,height=6cm,width=6cm}
\epsfig{file=t02.eps,height=6cm,width=6cm}

\caption{
Time evolution of \(f_n(x)\) (\(n = 0, 1, 2.\)) for 
\(r = 3.90, \epsilon = 0.80\).
The number of type-I and type-II fixed points increases with time.
Circles indicate fixed points, while arrows indicate
type-I fixed points, at which $f_n(x)$ intersects the identity function.
With the \(\epsilon f_n(x)\circ f_n(x)\) term, the function is distorted from
the simple iteration of the logistic map.}
\label{fig:evolve}
\end{figure}

\begin{figure}
\noindent
\epsfig{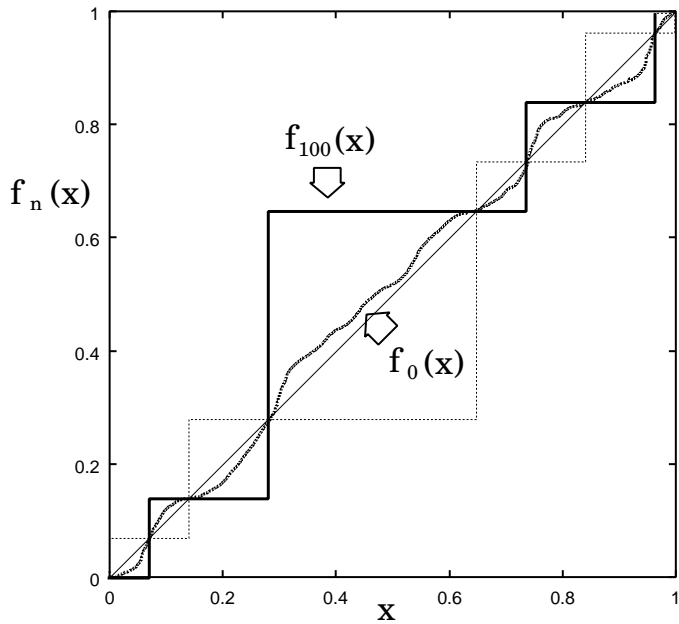}

\caption{For an initial continuous monotonically increasing function \(f_0(x)\),
the corresponding \(f_\infty (x)\) after the evolution is
plotted (here \(f_{100}(x)\) is plotted with \(f_0(x)\)). 
Once the initial function is given, the intervals in which $f_0(x) > x$ or $f_0(x) < x$ 
are determined as encircled by the dotted line.
In each interval, \(f_{\infty}(x)\) takes the same value as type-II fixed points.} 
\label{fig:mono}
\end{figure}

We can always arrange the \(M\) elements on a circle,
in a cyclic manner as $i$, $f(i)$, $f \circ f(i)$, $f \circ f \circ f(i)$$\cdots$.
Here we call this $f_0(i)$ a `cyclic network'.
The evolution of a cyclic network of $m$ elements is
shown in Fig.\ref{fig:e0peri} for $M=1,2,\ldots, 7$.

We can compute the period ($P(M)$) for each cyclic network.
In this case, each element $i$ connects to its left-hand neighbor
at first.  After the first iteration, the connection is changed to the
next nearest neighbor $i+2$.
With the next iteration, the connection is changed to $i+4$.  
In this way, after \(n\) iterations, the connection has changed to
$i + 2^n$. Since the element is mapped to
$2^n \pmod{m}$ neighbor at the $n$th step, it is convenient
to introduce the $a_n$ given by

\begin{equation}
 a_{n+1} = 2a_n \pmod{M}
\end{equation}
with \(a_0 = 1\).

When \(a_n\) reaches \(a_n = 1\), the \(M\) elements form 
a cycle of period \(n\). On the other hand, if \(a_n\) coincides with
one of previous $a_i$ \((i = 1,2,\ldots,n-1)\), the network
is reduced to a cycle with a period that is a divisor of \(M\).

As seen in Sec.4, a cyclic network consisting of an even number of elements is
reduced to disintegrated elementary networks, 
while a cyclic network consisting of an odd number elements is elementary.
By simple calculation, periods of some particular series can be obtained, such as
\(P(2^n -1) = n\) and \(P(2^n +1) = 2n\).
It is also found that there is a function \(\Phi (M)\) such that \(P(M) \le \Phi (M)\) 
from Euler's theorem. 
Here, the function $\Phi (M)$ (\(M\) is odd) is defined as follows.
Let us first decompose \(M\) into factors of prime numbers as
\(\prod_{k=1}^m {p_i}^{\alpha(i)}\), where \(p_i\) as a prime number.  Then
\(\Phi (M)\) is defined as the least common multiple among the values
\({p_i}^{\alpha(i)-1}(p_i -1)\) for \(i = 1,2,\ldots,m\).
\(\Phi (M)\) satisfies \(2^{\Phi (M)}\equiv 1 \pmod M\), but it is
not necessarily the smallest integer satisfying the condition. Hence,
the above inequality \(P(M) \le \Phi (M)\) is obtained.

Now let us consider the evolution of a network from a
general initial condition. If there are cyclic networks in the 
initial graph, the elements in the cyclic networks fall onto 
an elementary network, and they remain in this elementary network.
The cyclic network evolves only  within the elements
included in the initial network (i.e., $f^k(x)$ belongs to the network),
while the elements belonging to a line are eventually attracted to the network 
that the last element of the line is mapped to.

Now it is clear how the initial \(f_0 (i)\) is reduced to
a combination of elementary networks.
With the initial function, its configuration has been classified into 
(i) and (ii).  The final function $f_n(i)$ 
with $n \rightarrow \infty$ is given by
elementary networks (including type-I fixed points), and the
elements that are connected to them.
If the elementary network is a type-I fixed point, elements on the line
connected to it form type-II fixed points.\\

{\bf Acknowledgments}

The authors would like to thank Drs. T. Ikegami and S. Sasa for discussions.
This work is partially supported by Grant-in-Aids for Scientific Research from the Ministry of Education,
Science and Culture of Japan. 
One of authors (NK) is supported by a research fellowship from the Japan Society for the Promotion of Science.

\clearpage

\clearpage
\begin{figure}
\noindent
\epsfig{file=t1.eps,height=6cm,width=6cm}
\epsfig{file=t2.eps,height=6cm,width=6cm}
\end{figure}
\begin{figure}
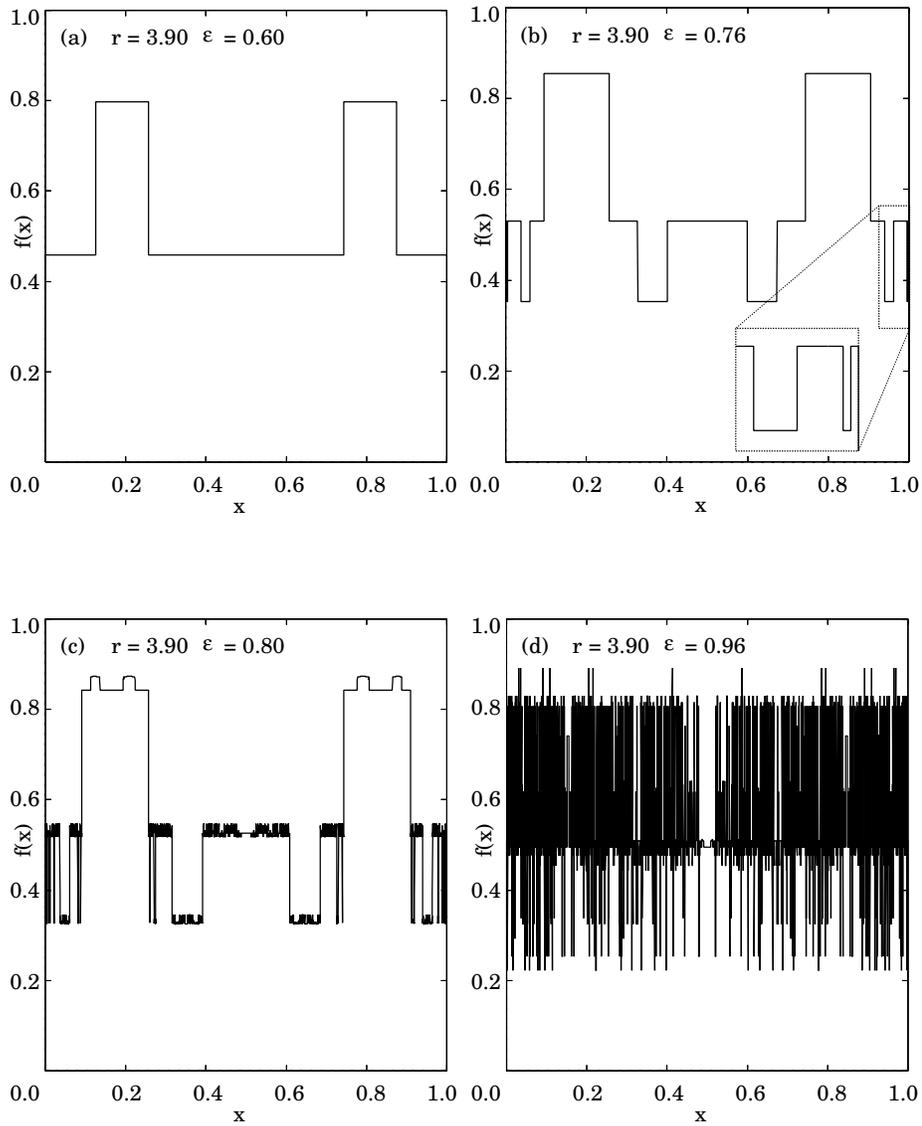

\noindent
\epsfig{file=t3.eps,height=6cm,width=6cm}
\epsfig{file=t4.eps,height=6cm,width=6cm}

\caption{
Four typical results of numerical simulations of our functional map,
with $M = 4096$, 
plotted at step 100, when $f(x)$ has converged to a fixed function.
(a) \(r = 3.9, \epsilon = 0.60\), (b) \(r = 3.9, \epsilon = 0.76\),
(c) \(r = 3.9, \epsilon = 0.80\), (d) \(r = 3.9, \epsilon = 0.96\).
By increasing \(r\) or \(\epsilon\),  discontinuous jumps spread all over the domain.}
\label{fig:type}
\end{figure}

\begin{figure}
\noindent
\epsfig{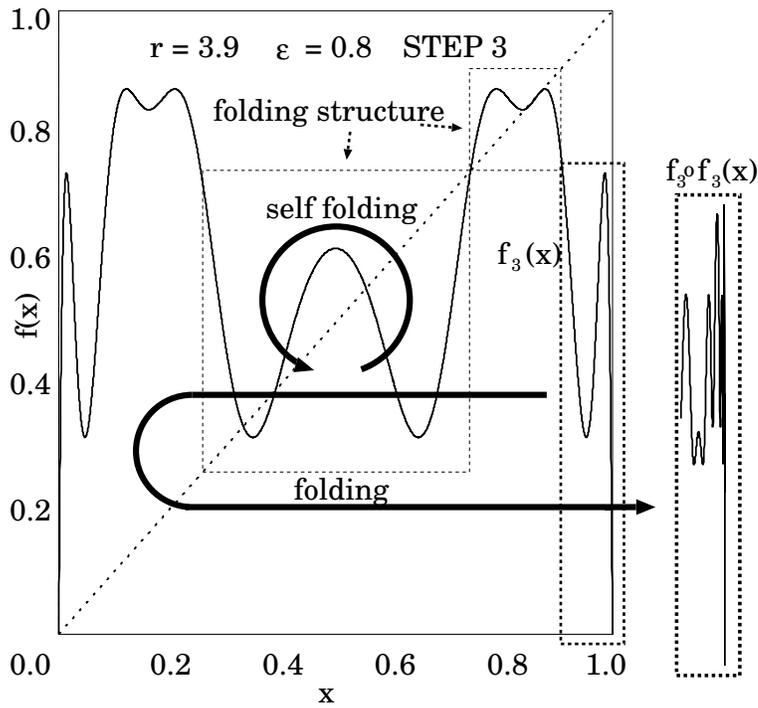}
\caption{The folding structure. 
A self-contained section (SCS) folds the region outside it and itself.
This is plotted at step 3, for \(r = 3.9\), \(\epsilon = 0.8\).
The dynamics of the region outside the SCS are driven by the structure in the 
SCS.
}
\label{fig:fff}
\end{figure}

\begin{figure}
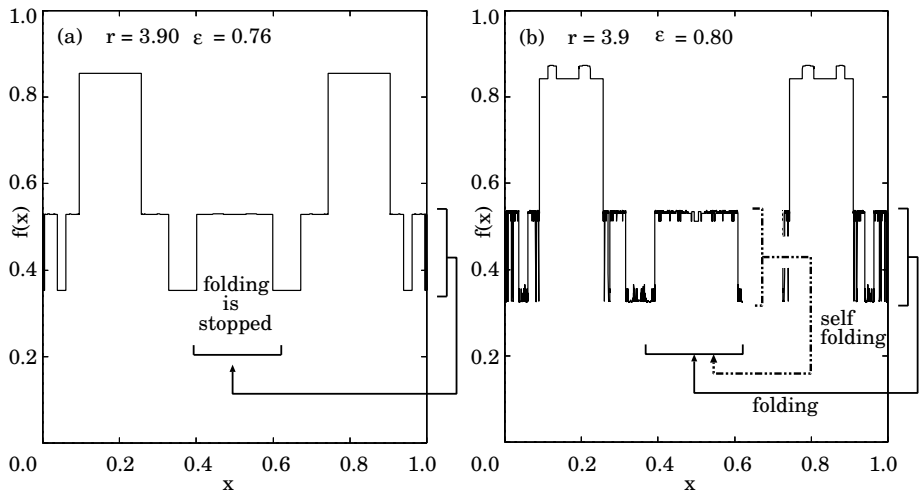

\noindent
\epsfig{file=fold1.eps,height=6cm,width=6cm}
\epsfig{file=fold2.eps,height=6cm,width=6cm}

\caption{ The folding mechanism in the SCS and the remaining part.
(a) Folding process at the center region stops to form steps, but 
it leads to fine step structures at
the edges of the map (near \(x\) = 0.0 and 1.0) during the
transient time, before the function (at the center) converges to a 
fixed function with few steps.
Here, \(r = 3.9, \epsilon = 0.76\).
(b) The center structure folds itself and the part outside the SCS.
This folding structure does not converge as \(n \rightarrow \infty\) and
forms finer and finer steps with time, leading to
fractal steps.
Here, \(r = 3.9, \epsilon = 0.80\).}
\label{fig:fold}
\end{figure}

\begin{figure}
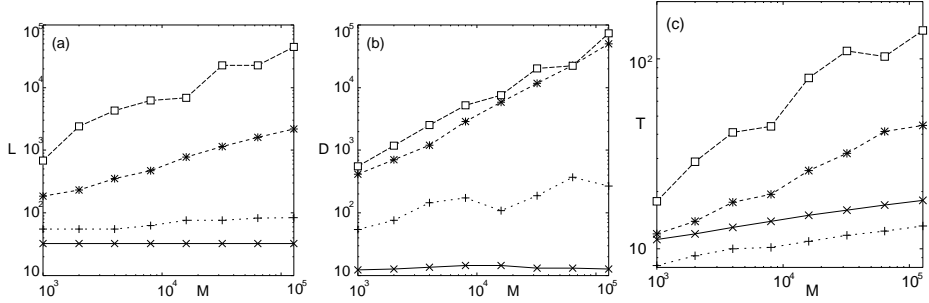

\noindent
\epsfig{file=len.eps,height=4cm,width=4cm}
\epsfig{file=dis.eps,height=4cm,width=4cm}
\epsfig{file=td.eps,height=4cm,width=4cm}

\caption{
Log-log plot of the mesh number $(M)$ dependence of
the $(L)$, $(D)$ and $(T)$: (a) the length of the graph $(L)$, 
(b) the discontinuous points $(D)$ and (c) the transient length $(T)$.
The largest mesh size is $128000$.
The length and discontinuous points are defined in the text, while the
transient is measured as the time steps before an initial
function is attracted to a fixed point for a given mesh.
four examples from the four phases are given, with
$r = 3.90, \epsilon = 0.60$ ($\times$);
$r = 3.90, \epsilon = 0.76$ ($+$);
$r = 3.90, \epsilon = 0.80$ ($*$);
and $r = 3.90, \epsilon = 0.96$ ($\Box$).
For (c), $\epsilon = 0.90$ is adopted for the plot with ($\Box$)
instead of 0.96.
}
\label{fig:quan}
\end{figure}

\begin{figure}
\noindent
\epsfig{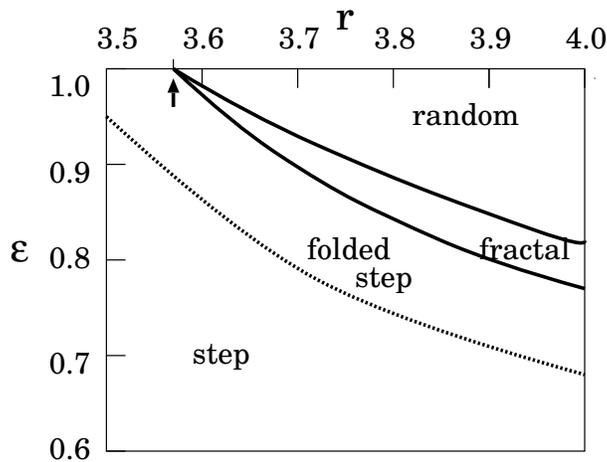}

\caption{
Phase diagram of our functional map with the initial logistic map
$rx(1-x)$. The phases (S), (FS), (F) and (R) are
classified according to the behavior of  \(L\), \(D\) and \(T\).
The point \(\epsilon = 1.0, r = 3.57\) corresponds to the onset of chaos of the logistic map.
}
\label{fig:phase}
\end{figure}

\begin{figure}

\caption{Bifurcation diagram.  Values of limiting forms of \(f(x')\) for all $x'$
(i.e., the values of all fixed points of $f_{\infty}(x)$)  are plotted
versus \(r\) with fixed \(\epsilon\) for all $x'$.
The values $f(x')$ for all $x'$ are
overlaid after the function converges under the iteration of the functional map.
(Thus, each point gives a value of a fixed point.)
(a) \(\epsilon = 0.7\), (b) \(\epsilon = 0.8\), (c) \(\epsilon = 0.9\).
Bifurcation collapse occurs at the parameter value where the points start to scatter.
This gives the onset of phase (F).}
\label{fig:bif}
\end{figure}

\begin{figure}

\caption{The function \(f_5(x)\) is plotted for (a) \(r = 3.80, \epsilon = 0.90\) 
and (b)\(r = 3.83, \epsilon = 0.90\).  They correspond to 
parameter values before and after the collapse of the lower bifurcation branch.
In (a), there remain 2 self-contained sections indicated by the squares.
In (b) the SCS at the lower value is collapsed, while the upper SCS
remains.}
\label{fig:dis}
\end{figure}

\begin{figure}
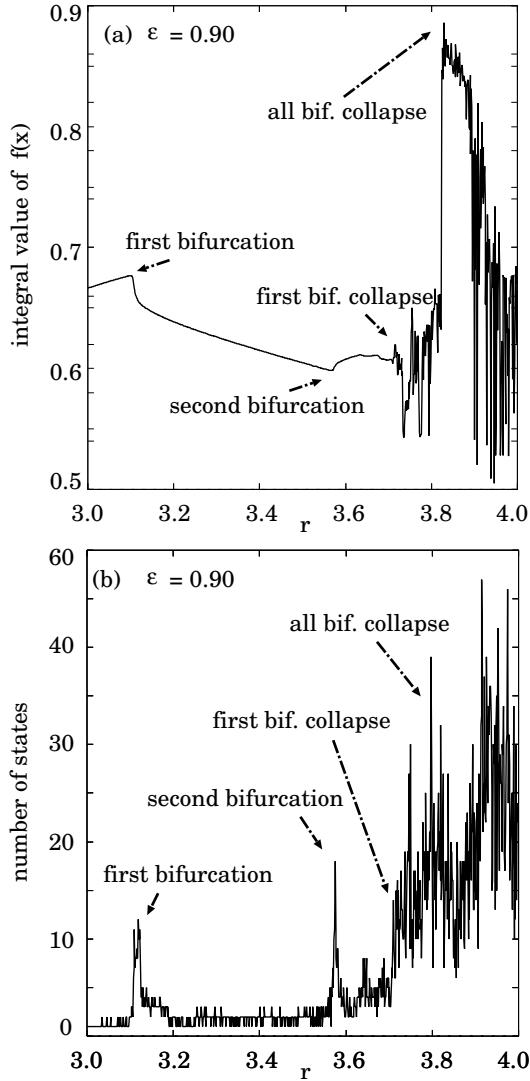

\noindent

\epsfig{file=sum.eps,height=6cm,width=7cm}
\epsfig{file=state.eps,height=6cm,width=7cm}

\caption{The parameter $r$ dependence of some characteristics.
(a) The quantity $\int dx f(x)$ versus \(r\) with fixed \(\epsilon\).
(b) The number of type-I fixed points versus \(r\) with fixed \(\epsilon\),
with mesh size 4096.
As mentioned, there are effects introduced by the finite size of the mesh:
addition of pseudo type-I fixed points near a tangent bifurcation, and 
missing type-I fixed points with a large slope in the random phase.
To remove the former effect, we count a series of 3 type-I fixed points
($x=(i-1)/M$, $i/M$, and $(i+1)/M$) as  one fixed point.
}
\label{fig:states}
\end{figure}

\begin{figure}
\noindent
\epsfig{file=peri1.eps,height=8cm,width=10cm}
\end{figure}
\clearpage

\begin{figure}
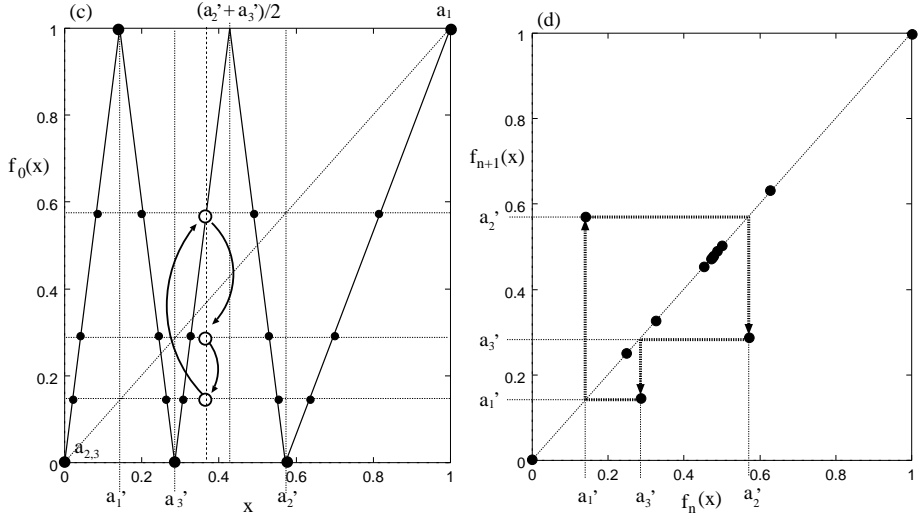

\noindent
\epsfig{file=p3.s.eps,height=6cm,width=6cm}
\epsfig{file=p333.eps,height=6cm,width=6cm}
\caption{
Periodic structures. 
(a) and (b) are schematic representations how at a given point $x = x'$ (a) or $x=x''$ (b) shows a period-2 motion.
In the representation,
$a_i$ is a type-I fixed point, while $a_i '$ is a type-II
fixed point mapped to it.  The value of $f_n(x')$ changes with period 2, where
$x'$ is mapped to $a_i'$.
In (b), the hierarchical configuration of periodic structure is shown,
where $f_n(x'')$ changes with period 2, mapping to the periodic points $x_i'$.
(c) An example of $f_0(x)$ giving a period-3 cycle. 
By starting from a function given by solid line,
$a_1 = 1$ and $a_2 ,a_3 = 0$ are type-I fixed points, and 
$a_1', a_2', a_3'$ are type-II fixed points mapped to $a_1, a_2, a_3$ respectively.
Then, the function values $f_n(x)$ in our model 
at small black points are mapped with period 3 as $a_1'$, $a_2'$ and
$a_3'$, successively.  As an example, the period 3 behavior is indicated by the 
white points and the arrows.
(d) A return map with starting $f_0(x)$ as (c) for all 
($f_n(x'), f_{n+1}(x')$) ($n = 1000 - 1010$) with $M = 8000$.
All $f_n(x')$ fall to a fixed point or a period-3 attractor.
In general, a map with \(k\) humps
has the possibility to form $k$-periodic points.
}
\label{fig:peri}
\end{figure}

\begin{figure}
\noindent
\epsfig{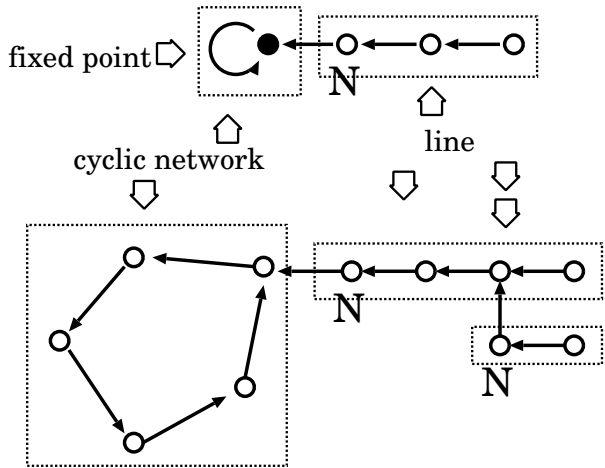}
\caption{Classification of the network.
Once an initial network is given,
it is separated into cyclic networks and lines.
}
\label{fig:e01}
\end{figure}

\begin{figure}
\noindent
\epsfig{file=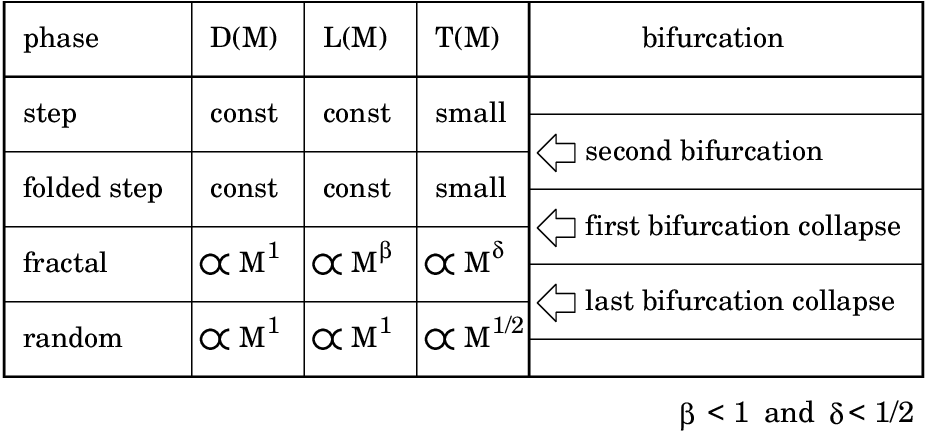,height=5cm,width=12cm}
\caption{Table 1: Characteristics of the four phases.
}
\end{figure}

\end{document}